\documentclass[sigconf]{acmart}

\usepackage{caption}
\usepackage{subcaption}
\usepackage{hyperref}
\usepackage{fancyref}
\usepackage{multirow}
\usepackage{array}
\usepackage{enumerate}
\usepackage{longtable}
\usepackage{colortbl}
\usepackage{csquotes}
\usepackage{booktabs}  
\usepackage{enumitem}
\usepackage{float}

\hypersetup{
    colorlinks=true,
    citecolor=[RGB]{0,112,125},
    linkcolor=blue,
    filecolor=red,      
    urlcolor=blue,
}

\colorlet{blue}{black}

\AtBeginDocument{%
  }

\copyrightyear{2026}
\acmYear{2026}
\setcopyright{cc}
\setcctype{by}
\acmConference[CHI '26]{Proceedings of the 2026 CHI Conference on Human Factors in Computing Systems}{April 13--17, 2026}{Barcelona, Spain}
\acmBooktitle{Proceedings of the 2026 CHI Conference on Human Factors in Computing Systems (CHI '26), April 13--17, 2026, Barcelona, Spain}
\acmPrice{}
\acmDOI{10.1145/3772318.3790816}
\acmISBN{979-8-4007-2278-3/2026/04}

\begin{document}

\title{Surfacing and Applying Meaning: Supporting Hermeneutical Autonomy for LGBTQ+ People in Taiwan}

\author{Yi-Tong Chen}
\orcid{0009-0000-3194-3659}
\affiliation{%
  \institution{National Taiwan University}
  \city{Taipei}
  \country{Taiwan}
}
\email{timicienio@ntu.im}

\author{En-Kai Chang}
\orcid{0009-0007-8060-7206}
\affiliation{%
  \institution{National Yang Ming Chiao Tung University}
  \city{Hsinchu}
  \country{Taiwan}
}
\email{debbychang.hk12@nycu.edu.tw}

\author{Nanyi Bi}
\orcid{0000-0003-1530-4464}
\affiliation{%
  \institution{National Taiwan University}
  \city{Taipei}
  \country{Taiwan}
}
\email{nanyibi@ntu.edu.tw}

\author{Nitesh Goyal}
\orcid{0000-0002-4666-1926}
\affiliation{
    \institution{Google Research}
    \city{New York}
    \country{United States}
}
\email{teshgoyal@acm.org}

\renewcommand{\shortauthors}{Chen et al.}

\begin{abstract}
After Taiwan’s legalization of same-sex marriage in 2019, LGBTQ+ communities continue to face hostility on social media. Using the lens of hermeneutical injustice and autonomy, we examine how technological conditions affect LGBTQ+ individuals’ identity exploration, narrative seeking, and community resilience. We conducted a multi-stage study with Taiwanese LGBTQ+ individuals, including in-depth interviews, participatory design workshops, and evaluation sessions. Participants described fragile yet creative strategies such as seeking validation in online interactions, reframing hostile content through theory, and relying on allies. Building on these insights, we designed and evaluated a retrieval-augmented, LLM-powered chatbot with four modes of interaction: reflection, validation, discussion, and allyship. Findings show that the system fosters hermeneutical autonomy by helping participants reframe hostile narratives, validate lived experiences, and scaffold identity exploration, while reducing the hermeneutical labor of navigating social media hostility. We conclude by outlining design implications for AI systems that advance hermeneutical autonomy through fluid self-representation, contextualized dialogue, and inclusive community participation.
\end{abstract}




%\received{04 September 2025}

\maketitle

\section{Introduction}
% \coffeestainB{0.9}{0.85}{-25}{5cm}{1.3cm}
% People, especially members of marginalized groups, rely on communities to interpret the world and their lived experiences \cite{TaylorBruckman-2024}. 

% 
Access to language, concepts, and frameworks necessary for interpreting lived experiences has been shown to be essential for the formation of identities and the daily navigation of social environments for LGBTQ+ people \cite{TaylorBruckman-2024}. Communities often provide them instrumental, social, and emotional support \cite{WangEtAl-2021b, CuiEtAl-2022, HaimsonEtAl-2020a, DymEtAl-2019}, which can come in the form of shared personal narratives, reflection on mutual trauma, articulation of concepts, and developing linguistic systems that make experiences intelligible \cite{Fricker-2007, TaylorBruckman-2024, AjmaniEtAl-2024}. Without such access, people may struggle to understand or communicate their experiences, leading to confusion, isolation, and internalized stigma \cite{AjmaniEtAl-2024}. In offline social situations where mainstream narratives leave marginalized people without supportive interpretive resources, online communities often serve to build collective concepts and vocabularies that counteract this systematic disadvantage \cite{HaimsonEtAl-2020, HardyEtAl-2022, CraigMcInroy-2014, SteedsEtAl-2025, TaylorBruckman-2024, ModiEtAl-2025, CuiEtAl-2022}. These shared interpretive and expressive resources have been referred to as \textbf{hermeneutical resources} as they enable communities to make meaning of complex situations \cite{Ferguson-2025, MilanoPrunkl-2025}. 

However, research has shown that the dynamics of online social networks can have limited benefit for the meaningful comparison of experiences for marginalized people \cite{MilanoEtAl-2021}. For example, members of the asexual community often suffer from a lack of understanding and awareness of their identity \cite{Lee-2021}. Despite multiple online resources that describe and explain this identity group\footnote{Such as: Taiwan Asexual Group (\url{https://www.facebook.com/asex.zh}) and Queer Margins (\url{https://www.queermargins.tw/-translations-ace-aro}).}, the Taiwanese general population continues to socially ostracize them due to familial values and an essentialist view of sexuality \cite{Lee-2021}. Since current resources are not integrated successfully in shaping an informed public opinion about real lived experiences, we wonder what kind of sociotechnical condition might be responsible for this. 
% Resources such as scientific studies, reports, and doctor-testimonials by American Medical Association helped shape informed public opinions in the West \footnote{https://www.ama-assn.org/public-health/population-health/advocating-lgbtq-community}.  
We further unpack that Taiwan provides a salient context for studying how LGBTQ+ people interact with such hermeneutical resources today.

% For instance, a victim of sexual harassment may lack the ``right words'' to name or describe the incident, excluding the experience from dominant constructions of a ``legitimate'' sexual offense. This exclusion silences individuals and perpetuates systemic injustice by reinforcing interpretive gaps \cite{Fricker-2007}. 

%%Yet many people move through the world without recognizing that these resources exist or that access to them varies. As a result, they may fail to leverage available resources for their well-being, reproducing hermeneutical injustice. This invisibility is often less consequential for non-marginalized groups, but can be critical for marginalized communities such as LGBTQ+ people \cite{Fricker-2007, TaylorBruckman-2024}.
% Against this backdrop Online communities and social media serve as a powerful epistemic infrastructure for the formation, uptake, and application of hermeneutical resources \cite{SafirEtAl-2025}. 

Although Taiwan became the first country in Asia to legalize same-sex marriage in May 2019 \cite{MinistryofJusticeTaiwan-2019}, acceptance from mainstream society remains uneven. Just six months earlier, the November 2018 referendum showed majorities opposed to LGBTQ-inclusive curricula and formal recognition of same-sex marriage in the Civil Code. Legalization also coincided with intensified everyday discrimination and mental health stressors \cite{Au-2022}. Among gay and bisexual young men, surveys reported high rates of harassment offline (60.3\%) and online (34.4\%) \cite{HuEtAl-2019}. Beyond the discourse on marriage, identities such as asexuality remain further marginalized \cite{Lee-2021}, while traditional Confucian family values continue to limit LGBTQ+ narratives \cite{AdamczykCheng-2015}. These dynamics push many LGBTQ+ people in Taiwan towards social networks to share experiences and craft collective narratives \cite{Lange-2015, Lee-2021}. 

Hence, we focus on Taiwan as a critical case in which progressive legal recognition (e.g., marriage equality in 2019) coexists with uneven social acceptance and Confucian-inflected family obligations. This legal–social mismatch and the importance of family harmony/collective norms make Taiwan uniquely qualified to observe how narratives are accessed and applied online when formal rights outpace cultural acceptance. {\color{blue} Further, global HCI and Queer HCI scholarship has historically privileged Western, “out-and-proud” models of queerness, obscuring how queerness is lived in contexts where privacy, relational obligation, and strategic invisibility are central survival practices \cite{ParkEtAl-2025, Kannabiran-2022, MoitraEtAl-2021, Taylor-2011}. Studying Taiwan, therefore, illuminates dynamics that are underrepresented in current HCI scholarship, responds directly to calls for greater recognition of non-Western epistemologies, local knowledge, and unique contextual specificity in HCI \cite{ParkEtAl-2025, Taylor-2011}}. We therefore ask:
 
\textbf{RQ1: How do members of Taiwan's LGBTQ+ group manage hermeneutical resources on social media?}

{\color{blue}
Next, we argue that we need a systematic mechanism to support the creation, sharing, and usage of such hermeneutical resources. Traditionally, institutions like schools, universities, and libraries serve as sites where people acquire these interpretive resources. In recent years, social media has emerged as a powerful and dynamic channel for marginalized groups to share and exchange their experiences \cite{MilanoPrunkl-2025}. For instance, Taylor \& Bruckman \cite{TaylorBruckman-2024} described the lively and creative ways of the \textit{r/bisexual} subreddit members engage with the discourse on bisexuality through collective construction of terms, jokes, stories, and memes. This creative engagement helped both questioning and established members make sense of sexuality-related issues. In this case, a corner of social media effectively served as an \textbf{epistemic infrastructure} for a marginalized community, which enabled the \textit{formation}, \textit{uptake}, and \textit{application} of hermeneutical resources that countered wider societal misconceptions about bisexuality \cite{MilanoPrunkl-2025}. 

We seek to extend this formulation by exploring how marginalized communities can be empowered to undertake such infrastructure work across the wider social media ecosystem and diverse social contexts. Specifically, we investigate how hermeneutical resources can \textit{travel with} the user as they navigate the social sphere, allowing for contextual access and application in various social situations. Members of the LGBTQ+ community and beyond are increasingly using LLM-powered chatbots to make sense of social experiences and to rehearse social behaviors \cite{WangEtAl-2021b, HuangEtAl-2024a, LiuEtAl-2025}, positioning these systems as convenient sources of knowledge that can complement online communities. We wonder if such technologies might be helpful in creating the infrastructure needed.

However, leveraging LLMs can be challenging for culturally non-Western users, as current models are trained predominantly on Western datasets and may encode bias \cite{GwagwaMollema-2024, Katz-2020, MuldoonWu-2023}. For example, algorithmic targeting \cite{Stewart-2022}, prejudice \cite{OriggiCiranna-2017}, and over-personalization \cite{MilanoEtAl-2021} with limited data on marginalized non-Western users can distort what they see on social networks. So, instead of discovering supportive hermeneutic resources on such platforms, they may find themselves further isolated. %%For example, if a person is shown only advertisements for low-paying jobs, it becomes difficult to discern whether this is commonplace or discriminatory. 

%%For Taiwanese LGBTQ+ people, this adds another layer of complexity to using LLMs for accessing adequate support. 

All hope is not lost. Technological practices, such as retrieval augmented generation (RAG) and the understanding of users' natural language, are beginning to reveal possibilities for making conversational agents more culturally sensitive and understandable to users \cite{ChangEtAl-2024, SeoEtAl-2025, PetridisEtAl-2024, YuanEtAl-2025, ZhaoEtAl-2024, HuangEtAl-2024a, RadlinskiEtAl-2022, ChenEtAl-2025}. This points to a possibility for AI systems to be more easily designed by and for marginalized communities, even without the advancement of mainstream LLMs. Accordingly, we ask:}

\textbf{RQ2: How can technologies such as LLM-powered agents support the uptake and application functions of epistemic infrastructure for LGBTQ+ people?}

{\color{blue}To investigate these questions, we conducted a multi-stage study combining formative interviews, design workshops, and prototype evaluation. In the formative phase, we interviewed 10 Taiwanese LGBTQ+ individuals about their social media experiences, focusing on how they accessed and applied hermeneutical resources in moments of stigma or misunderstanding. Based on the findings from our formative study, we foreground \textit{hermeneutical injustice} \cite{Fricker-2007, Medina-2021, TaylorBruckman-2024} as a critical lens to analyze the present unfair epistemic condition faced by Taiwanese LGBTQ+ people on social media, and inform our subsequent development of a supportive prototype.

We then ran design workshops with 8 participants to co-create and critique Queerbot, an LLM-powered agent designed to scaffold \textit{hermeneutical autonomy} \cite{AjmaniEtAl-2025} by helping participants retrieve relevant resources in context, compare experiences, and craft more intelligible responses. Importantly, participants envisioned AI as a collaborator, rather than a proxy, that allows users to interpret and testify to their lived experiences.

Finally, in three evaluation sessions, 11 participants interacted with the prototype and reflected on its capacity to support identity expression, interpretive empowerment, and discourse participation. Our findings show that Taiwanese LGBTQ+ people frequently encounter hermeneutical injustice online, particularly when supportive narratives are obscured by algorithms or when they lack confidence in making their experiences intelligible to others.}

This work builds upon prior research on LGBTQ+ social media experiences and clarifies how platforms can either shape or undermine marginalized individuals' capacity to interpret and communicate their lives. Our work bridges the \textit{formation} of hermeneutical resources within LGBTQ+ communities and the \textit{access} and \textit{application} of these resources in everyday social media interactions. By designing an agent capable of retrieving, contextualizing, and supporting the use of such resources, we demonstrate how community-driven knowledge can be operationalized to reduce hermeneutical labor and foster hermeneutical autonomy outside safe spaces.

\section{Related Work}
In this section, we review scholarship on hermeneutical injustice and autonomy, examine how social media functions as epistemic infrastructure in LGBTQ+ contexts with a focus on Taiwan, and discuss emerging work on social support AI and LLMs as potential extensions of these infrastructures.  

\subsection{Hermeneutical Injustice and Hermeneutical Autonomy on Social Media} 
In our formative study, as described in Section 3.3, we discovered that many people move through the world without recognizing that helpful resources exist or that access to them varies. As a result, they may fail to leverage available resources for their well-being, reproducing hermeneutical injustice. This invisibility is often less consequential for non-marginalized groups, but can be critical for marginalized communities such as LGBTQ+ people \cite{Fricker-2007, TaylorBruckman-2024}.
% Against this backdrop Online communities and social media serve as a powerful epistemic infrastructure for the formation, uptake, and application of hermeneutical resources \cite{SafirEtAl-2025}. 

Access to appropriate language, concepts, and frameworks is critical for interpreting lived experiences, shaping identity, and contributing to inclusive public discourse. These interpretive toolkits, or \textbf{hermeneutical resources}, emerge through community building and the sharing of experiences \cite{AjmaniEtAl-2025}. 
Accordingly, Fricker \cite{Fricker-2007} describes \textit{hermeneutical injustice}, a form of \textit{epistemic injustice}, as occurring when marginalized groups are systematically excluded from the \textit{formation}, \textit{access}, or \textit{application} of hermeneutical resources necessary to articulate their lives \cite{Fricker-2007, MilanoPrunkl-2025}. 
At the individual level, such exclusion undermines the ability to make sense of and confidence in articulating one’s experiences, furthering epistemic marginalization. At the collective level, it prevents groups from shaping shared vocabularies and interpretations, thereby limiting their ability to influence broader social discourse \cite{Fricker-2007, GeorgeGoguen-2021}.

{\color{blue} Mitigating such injustice necessitates \textit{hermeneutical autonomy}, or the ability to leverage resources necessary to describe one's lived experience \cite{AjmaniEtAl-2025}.} Online LGBTQ+ communities demonstrate this autonomy by creating and circulating community-specific vocabularies that validate identities on their own terms \cite{TaylorBruckman-2024, AjmaniEtAl-2023, AjmaniEtAl-2024, DymEtAl-2019, FoxRalston-2016, SimpsonSemaan-2021}. For example, research on the \textit{r/bisexual} subreddit shows how the absence of bisexual knowledge in mainstream discourse produced hermeneutical injustice, leaving bisexual individuals misrecognized by others. In response, the community developed \textit{cultural toolkits}, which are language, jokes, memes, and practices that resisted erasure and articulated both the being and doing of bisexuality \cite{TaylorBruckman-2024, AjmaniEtAl-2024, AjmaniEtAl-2023}. Such vocabularies enable marginalized identities to become intelligible outside dominant hetero- and monosexual norms.

\begin{figure}[t!]
    \centering
    \includegraphics[width=0.99\linewidth]{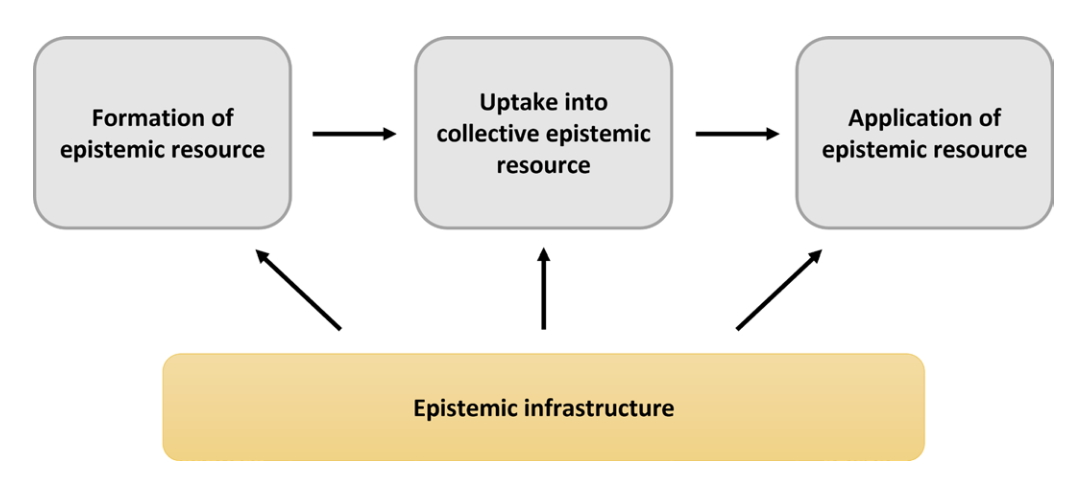}
    \caption{The role of epistemic infrastructure in supporting the sharing and comparing of experiences, enabling the formation, uptake, and application of interpretive resources, as identified by Milano \& Prunkl \cite{MilanoPrunkl-2025}.}
    \Description{Diagram illustrating the support of epistemic infrastructure for three resource stages. A wide bar at the bottom labeled 'Epistemic infrastructure' points upward with three separate arrows to a horizontal sequence of three boxes above. The top sequence flows via arrows from 'Formation of epistemic resource' to 'Uptake into collective epistemic resource' and finally to 'Application of epistemic resource'.}
    \label{fig:epistemicInfra}
\end{figure}

Building on this foundation, Milano \& Prunkl \cite{MilanoPrunkl-2025} conceptualize social media as a form of \textbf{epistemic infrastructure}, which mediates the \textit{formation} (accumulation and exchange), \textit{uptake} (access and spread), and \textit{application} (usage across contexts) of epistemic resources (see fig. \ref{fig:epistemicInfra}). Social media can therefore support hermeneutical autonomy by enabling marginalized groups to connect with peers, share experiences, and mobilize interpretive resources. 

At the same time, the failures of epistemic infrastructure can deepen hermeneutical injustice. Research highlights how algorithmic sorting and targeting \cite{Stewart-2022}, identity prejudice \cite{OriggiCiranna-2017}, and \textit{epistemic fragmentation} through over-personalization \cite{MilanoEtAl-2021} isolate individuals and hinder the meaningful sharing and comparison of experiences. These dynamics restrict marginalized people’s ability to form, access, and advocate for hermeneutical resources, perpetuating unfair epistemic conditions \cite{Stewart-2022}. 
{\color{blue} Building upon the theoretical groundwork established by Ajmani et al. \cite{AjmaniEtAl-2025} on the governance of hermeneutical resources, we extend this lens to examine how hermeneutical autonomy is further shaped and at times constrained by the epistemic infrastructures that mediate the \textit{uptake} and \textit{application} of these interpretive resources.}

Moreover, because social media platforms are embedded within cultural and societal contexts, their use is shaped by dominant norms reinforced through community practices \cite{AbiriBuchheim-2022}. Thus, how individuals leverage hermeneutical resources to achieve autonomy is conditioned not only by platform affordances but also by the cultural backgrounds of their users. In the following section, we situate our study within the specific cultural and social context of Taiwan.

\subsection{Taiwanese LGBTQ+ Culture on Social Media} 
As a historically marginalized group, LGBTQ+ communities’ engagement with social media has been a central focus of HCI research \cite{TaylorEtAl-2024a}. Prior studies have examined how LGBTQ+ individuals manage \cite{CraigMcInroy-2014, Cavalcante-2016, CarrascoKerne-2018, Haimson-2018, DymEtAl-2019, SteedsEtAl-2025} and explore their identities \cite{Wuest-2014, HaimsonEtAl-2020}, as well as how social media facilitates community support and resource sharing \cite{HaimsonEtAl-2020, HardyEtAl-2022, CraigMcInroy-2014, SteedsEtAl-2025, TaylorBruckman-2024, ModiEtAl-2025, CuiEtAl-2022}. Central to these practices is the creation of ``safe spaces'' that balance opportunities for connection with privacy concerns \cite{Duguay-2016, CraigEtAl-2020, ZhuSkoric-2021, ZhuSkoric-2025}. Social media has also been linked to fostering resilience among LGBTQ+ youth, who often face isolation and limited support offline \cite{CraigEtAl-2015, CraigEtAl-2020}.

Yet research also identifies tensions between LGBTQ+ community needs and the affordances of mainstream social platforms \cite{ArmstrongEtAl-2024, HardyEtAl-2022, HaimsonEtAl-2020, HaimsonEtAl-2020a, BivensHaimson-2016, DymEtAl-2019, HardyVargas-2019}. Embedded design values may conflict with LGBTQ+ practices \cite{DeVitoEtAl-2021a}, while intra-community dynamics, such as peer pressure or intersectional marginalization (e.g., transgender people of color), complicate platform use \cite{StarksEtAl-2019, ModiEtAl-2025, HaimsonEtAl-2021, WalkerDeVito-2020}. In response, scholarship emphasizes surfacing values rooted in LGBTQ+ lived experiences, including meaningful connection \cite{ArmstrongEtAl-2024, HaimsonEtAl-2020}, discovery of safe or affirming spaces \cite{HardyEtAl-2022, HardyVargas-2019}, and envisioning dedicated LGBTQ+ platforms \cite{HaimsonEtAl-2020a}.

Cultural values and heritage also shape social media practices. Orientations such as individualism–collectivism and long-term orientation influence motives for sharing, privacy concerns, and sustained engagement \cite{SharminEtAl-2021}. Platform adoption and content circulation similarly reflect cultural orientations; for example, individualist settings show higher levels of cross-cultural video consumption and knowledge sharing \cite{ParkEtAl-2017}. In heritage-related exchanges, communities rely on narratives, rituals, and trusted authorities as credibility cues, while digitization and immersive media raise concerns over authenticity and visibility \cite{BonacchiEtAl-2023, TaylorGibson-2017, WenEtAl-2025}. These dynamics highlight how cultural heritage governs not only interpretation but also participation in online environments.

Taiwan’s legalization of same-sex marriage in 2019 \cite{MinistryofJusticeTaiwan-2019} was a landmark event, but enduring ambivalence reflects tensions between progressive reforms and traditional values. Confucian-informed norms emphasizing family continuity are linked to lower tolerance of homosexuality in East Asian societies \cite{Adamczyk-2017, Au-2022}. At the same time, urbanization and new media have enabled adolescents greater autonomy in everyday decision-making \cite{HsiehEtAl-2004}. Yeh’s concept of \textit{reciprocal filial piety} further illustrates how autonomy and familial obligations can coexist through mutual care and respect \cite{Yeh-2003, Yeh-2014}. This framework helps explain how LGBTQ+ youth may negotiate self-exploration while maintaining family harmony. 

Against this cultural backdrop, online communities play a vital role for Taiwanese LGBTQ+ people. Digital platforms provide linguistic and cultural resources for self-articulation \cite{Lee-2021} and serve as crucial venues for empowerment and belonging, enabling resistance to offline marginalization \cite{Lange-2015}. Building on this context, our study foregrounds \textit{hermeneutical injustice} as a lens for examining Taiwanese LGBTQ+ social media practices, linking macro-level cultural tensions with micro-level struggles of identity interpretation. However, to the best of our knowledge, no studies directly investigate how Taiwanese LGBTQ+ individuals leverage hermeneutical resources on social media. Therefore, we ask:  

\textbf{RQ1: How do the Taiwanese LGBTQ+ group members manage the hermeneutical resources on social media?}

\subsection{Designing Social Support AI to Foster Hermeneutical Autonomy}
Conversational agents are increasingly adopted for emphatic, context-aware social companionship \cite{ChaturvediEtAl-2023, VermaEtAl-2025}. Within this trend, research on social support chatbots has gained momentum \cite{WangEtAl-2021b, LatifEtAl-2024, ChaturvediEtAl-2023, LatifEtAl-2024}. These systems provide timely support and facilitate exchanges, proving especially effective in health contexts where community-curated knowledge addresses both \textit{informational} and \textit{emotional} needs \cite{WangEtAl-2021b}. The rise of large language models (LLMs) in recent years has further advanced these capabilities, offering accessible, empathetic companionship \cite{LiuEtAl-2025a, LiuSundar-2024, VermaEtAl-2025, TaEtAl-2020}, {\color{blue}as well as aiding social interactions \cite{FuEtAl-2024, FuEtAl-2025, HohensteinEtAl-2023}}. 

{\color{blue} Despite its potential, however, deploying AI to mediate social interactions introduces significant challenges. Research has shown that when acting as a communication aid, AI tends to amplify emotions \cite{FuEtAl-2024}, yield context-inappropriate outputs \cite{FuEtAl-2024, FuEtAl-2025}, and is often perceived as inauthentic \cite{JakeschEtAl-2019}. In the long term, relying on AI for social support risks eroding user autonomy \cite{FuEtAl-2025, ZhangEtAl-2025} and fostering emotional dependence \cite{ManziniEtAl-2024, ZhangEtAl-2025}. These harms are often driven by AI model's inherent tendency toward agreement with prompters \cite{FuEtAl-2025} and the commercial incentives of platforms to maximize engagement \cite{ZhangEtAl-2025, KirkEtAl-2025}. Scholars have also warned that AI risk reproducing existing stigma, bias, and epistemic injustice associated with social factors \cite{MarkoEtAl-2025, KayEtAl-2025, MeiEtAl-2023, BragazziEtAl-2023, PetzelSowerby-2025, ByrnesSpear-2023}, including normative assumptions of gender and sexual orientation \cite{TaylorEtAl-2025, ScheuermanEtAl-2019}, as authentic representations of localized, community-centered voices remain scarce in the development of AI systems \cite{KayEtAl-2025, QadriEtAl-2023}. }

% These concerns underscores the need for culturally inclusive and community-driven design that aligns system affordances with the values and practices of traditionally marginalized groups \cite{WangEtAl-2021b, KayEtAl-2025, WanEtAl-2023}.}

% These implications are even more relevant when concerning traditionally marginalized and stigmatized populations such as LGBTQ+ people \cite{BragazziEtAl-2023, PetzelSowerby-2025a}.
% Kirk et al. \cite{KirkEtAl-2025} identified these dynamics as \textit{social reward hacking}, where systems exploit human social reward mechanism through tactics like flattery or emotional appeals to optimize for short-term satisfaction rather than long-term wellbeing. Consequently, they advocated for a shift of social AI design towards \textit{socioaffective alignment}, which specifically urges designers to resolve intrapersonal dilemmas such as the conflict between immediate gratification and long-term growth, in order to support and align with the user's basic psychological needs for autonomy, competence, and relatedness \cite{KirkEtAl-2025, RyanDeci-2008, RyanSapp-2007}.}

{\color{blue}Recent work addresses these concerns through two LLM-related techniques: retrieval-augmented generation (RAG) and natural language user knowledge. RAG supplements model parameters by looking up external knowledge bases at inference time, which improves factual accuracy, contextual relevance, and cultural sensitivity \cite{ChangEtAl-2024, SeoEtAl-2025, SultanaEtAl-2025}. Encoding system's knowledge of user personas \cite{HuangEtAl-2024a, RadlinskiEtAl-2022} and preferences \cite{PetridisEtAl-2024, YuanEtAl-2025, ZhaoEtAl-2024} in natural language offers personalization with transparency and control, allowing individuals to see and shape how systems understand them, reclaiming (part of) the autonomy when engaging and aligning with AI agents  \cite{HuangEtAl-2024a, RadlinskiEtAl-2022, ChenEtAl-2025}. These approaches suggest promising directions for mitigating algorithmic hermeneutical injustice and fostering responsive support for underrepresented groups \cite{KayEtAl-2025}, while preserving basic human needs of self-determination \cite{KirkEtAl-2025, RyanDeci-2008, RyanSapp-2007}.} 
%Building on these foundations, in this study, we design and prototype an AI system capable of retrieving, contextualizing, and applying hermeneutical resources created within LGBTQ+ communities in a socially contextualized manner. 
However, to the best of our knowledge, no prior work has examined how such social support systems can afford the \textit{uptake} and \textit{application} functions of epistemic infrastructure for culturally marginalized groups in a social setting. Therefore, we ask:

\textbf{RQ2: How can technologies such as LLM-powered agents support the uptake and application functions of epistemic infrastructure for LGBTQ+ people?}

{\color{blue}

%In order to investigate these issues, we conducted a three-stage study involving formative interviews, design workshops, and prototype evaluation workshops. In the spirit of the participatory design (PD) research tradition \cite{SimonsenRobertson-2012}, our work collaborated with LGBTQ+ community members through co-creative design and evaluation activities, which were intended to empower community members to meaningfully shape a sociotechnical system important to their lives and to bring their lived experiences and context-embedded knowledge into academic discussion. 
To empower community members to meaningfully shape a sociotechnical system important to their lives and to bring their lived experiences and context-embedded knowledge into the design process and, by extension, academic discussion, we collaborated with LGBTQ+ community members through co-creative design and evaluation activities, in the spirit of the participatory design (PD) research tradition \cite{SimonsenRobertson-2012}. Our work also echoes later iterations of value-sensitive design (VSD) \cite{Friedman-1996, LeDantecEtAl-2009, BorningMuller-2012}, which called for respecting local values based on direct investigation of community members’ lived experience \cite{LeDantecEtAl-2009, BorningMuller-2012, DeVitoEtAl-2021a}. 
To answer the research questions, we organized the study into three progressive phases involving formative interviews, design workshops, and prototype evaluation, which are presented in the sections that follow.
% In our study, we critiqued the epistemic conditions currently faced by Taiwanese LGBTQ+ people on social media through the lens of hermeneutical injustice. This allowed us to identify community members’ core value of hermeneutical autonomy and consider how it should guide the prototype’s features and system architecture.

}

\section{Formative Study}
The formative study aims to understand Taiwanese LGBTQ+ users' social media experience though the lens of hermeneutical injustice and autonomy.
Specifically, we aim to understand how the availability of hermeneutical resources supports Taiwanese LGBTQ+ in interpretation and responding when they face sociopolitical stressors on social media.
In this section, we report the methodological details and results of the formative study.
\begin{table*}[t]
  \caption{Demographic information of the formative study participants}
  \label{tab:formative-demo}
  \Description{Table displaying demographic data for ten participants (P1 to P10). Columns include Participant ID, Age, Occupation, Gender, and Sexual Orientation. Ages range from 21 to 24. Occupations include eight students, one teacher (P3), and one developer (P4). Gender distribution consists of four Cis Males, four Cis Females, and two Non-binary individuals. Sexual orientations listed are Homosexual (six participants), Pansexual (two participants), Androsexual (one participant), and Asexual (one participant). Superscripts a through d mark specific terms in the Gender and Sexual Orientation columns.}
  
  % Minipage ensures footnotes stay with the table and use alphabetic markers
  \begin{minipage}{\textwidth}
    \centering
    \begin{tabular}{cclll}
      \toprule
      \textbf{Participant ID} & \textbf{Age} & \textbf{Occupation} & \textbf{Gender} & \textbf{Sexual Orientation} \\
      \midrule
      P1  & 24 & Student   & Cis Male   & Homosexual \\
      P2  & 23 & Student   & Cis Male   & Homosexual \\
      P3  & 23 & Teacher   & Cis Male   & Homosexual \\
      P4  & 24 & Developer & Cis Female & Homosexual \\
      P5  & 22 & Student   & Non-binary\footnote{Describing a person who experiences their gender identity and/or gender expression as falling outside the binary gender categories of man and woman \cite{GLAAD-2022}.} 
                                        & Androsexual\footnote{Describing a person who is primarily sexually, aesthetically, and/or romantically attracted to masculinity \cite{GLAAD-2022}.} \\
      P6  & 21 & Student   & Cis Female & Homosexual \\
      P7  & 23 & Student   & Cis Female & Homosexual \\
      P8  & 24 & Student   & Non-binary & Asexual\footnote{Describing a person who does not experience sexual attraction \cite{GLAAD-2022}.} \\
      P9  & 22 & Student   & Cis Male   & Pansexual\footnote{Describing a person who has the capacity to form enduring physical, romantic, and/or emotional attractions to any person, regardless of gender identity \cite{GLAAD-2022}.} \\
      P10 & 23 & Student   & Cis Female & Pansexual \\ 
      \bottomrule
    \end{tabular}
  \end{minipage}
\end{table*}

\subsection{Method}
We conducted face-to-face in-depth interviews with 10 participants of various LGBTQ+ identities during a two-week period in March 2025. 
This study, as well as the following two studies in the paper, have received IRB approval from National Taiwan University, with the protocol ID \#202503HS019.
In this subsection, we report the protocol, participants, and procedure of the interviews.

\subsubsection{Interview protocol}
The interview protocol is structured around {\color{blue}three} thematic areas that directly address the study's focus on the participation of Taiwanese LGBTQ+ users leveraging hermeneutical resources in interpreting and responding to sociopolitical stressors on social media.

\paragraph{Basic information.}
This section collects demographic and identity-related data (age, occupation, gender identity, sexual orientation) to contextualize the experiences of the participants and situate their narratives within relevant social and cultural backgrounds.

\paragraph{Social media practices.}
These questions map the social media usage patterns, the preferences of the platform, and the participation of the community, including the participation of spaces related to LGBTQ+ and non-LGBTQ+. This information helps identify the digital contexts in which sociopolitical stress occurs and where hermeneutical injustice comes into play.

\paragraph{Stress, and interpretation and response.}
This section elicits in-depth accounts of encountering online and offline negative or hostile LGBTQ+-related content, as well as their interpretation of, and response to the situation (e.g., seeking support, shifting focus, fighting back).  
In addition, it explores the influence of large-scale sociopolitical events, such as the 2018 Taiwan same-sex marriage referendum. These narratives provide rich qualitative data to understand how sociopolitical stressors manifest in digital environments and how participants perceive the role of hermeneutical resource in the process. The complete content of the interview protocol is detailed in Appendix \#1.

% \paragraph{Self-reflection behaviors and design considerations.}
% This final theme examines the self-assessment of the participants of their strategies for using social networks and regulating emotions, perceived growth in psychological resilience, and recommendations for features of the platform (e.g., usage pattern analysis, mental health prompts) that could better support emotional well-being.

Through this structure, the protocol captures (1) the situational triggers and contexts of sociopolitical stress for Taiwanese LGBTQ+ users on social media, (2) the affordances and hermeneutical resources these users appropriate to navigate such situations, which in turn cultivate resilience.

\subsubsection{Participants}

10 Participants were recruited using the snowball method, primarily through connections of the first author (hereafter referred to as FA). A majority of the participants are cisgender individuals (N = 8), of whom most identify as homosexual (N = 6), while others identify as pansexual (N = 2). Of the two participants who identified as non-binary, one identified as androsexual and the other as asexual. Their complete demographic information was shown in table \ref{tab:formative-demo}.

\subsubsection{Procedure}

We conducted one-on-one in-person interviews with each of the participants. Each interview lasts for approximately 60 minutes. With consent from the participant, we audio-recorded every interview. After the interviews, each participants were compensated with NT\$200 (about 6.6 USD) for their participation. 

\subsection{Data analysis}
{\color{blue}
After transcribing each interview recording into text, we performed inductive thematic analysis \cite{BraunClarke-2006, TaylorBruckman-2024} on our data. Specifically, after familiarizing ourselves with the transcripts, the first author performed the initial, open coding of the dataset, resulting in a preliminary set of 231 codes. To enhance credibility and ensure analytic rigor, the initial codes and their application were reviewed collaboratively by the research team to develop the final coding scheme. In instances of disagreement, all authors engaged in extensive discussion to reach a full consensus on the final code application and structure. 

Codes were then further grouped and labeled into themes using affinity diagrams on Figma, each reflects common practices, desires, and challenges among the participants. The themes were then discussed, refined, and confirmed during meetings among the authors. During these meetings, we noticed how participants consistently sought out and utilized various forms of informational and social support to interpret and respond to the sociopolitical stressors they encountered online. We also identified moments where the participants struggled to find adequate language, validating narratives, or supportive communal feedback. These observations led us to adopt \textit{hermeneutical injustice} \cite{Fricker-2007} as our primary theoretical lens, both for interpreting data from the formative study, as well as inferring the latent values of our participants that guide subsequent studies.}

\subsection{Findings}

\subsubsection{Situating own identity and life experiences with narratives}

% While our participants usually first noticed their experiences being different from the cisgender/heterosexual norm in early stages of life, individuals later acquired the term that describes, or the supporting narrative that validates their experiences. This process often involves putting oneself in a predefined label or category, and a period of adjusting to the corresponding expectations of the identity. The result could be the individual embracing (at least temporarily) the identity, leaving it in favor of a more suitable one, or rejecting labels altogether.

Online communities were a common avenue through which participants came to recognize and articulate their identities. By engaging with community-specific narratives and interacting with other members, they accessed resources that helped them make sense of their own experiences. The anonymity and exclusivity of these spaces made them particularly appealing for exploring intimate personal issues. For instance, P6 recalled noticing affectionate feelings toward female classmates as early as elementary school but lacking the vocabulary to describe them until she encountered lesbian communities online: 

\begin{quotation}
\textit{``There was kind of a trend to classify lesbians into different types. And that’s when I realized—oh, I could probably be categorized into one of those groups... Back then, there were already anonymous lesbian chat apps, and everyone there would start by asking, ‘Which category are you?’ So gradually, that really deepened my knowledge of and identification with that label.''} (P6)
\end{quotation}

Some participants gained identity-related knowledge through traditional media or professional narratives, such as books and articles, which offered relatively comprehensive and structured accounts of insights about identities. These kinds of more in-depth hermeneutical resources, often accessed when participants attained higher levels of mental maturity and narrative literacy, helped individuals make sense of their lives in a broader way than personal stories shared in online communities. P9, who once identified as gay, described his ``nerdy'' approach to re-examining his sexual orientation after entering a relationship with a trans man, drawing on online materials about alternative identities and articles from credible sources:

\begin{quotation}
\textit{I think it was only after I started dating a transgender partner that I really came to understand gender in a different way, with more perspectives. After that, I looked into what exactly `gay,' `bisexual,' and `pansexual' mean. Beyond that, the kinds of articles I read were usually from sources with more credibility—like psychologists explaining how to get through these challenges. Of course, there are a lot of online forums, like Reddit, where people share things... I think that’s helpful too, but most of that is personal experience. And I feel like you can’t necessarily apply those directly to everyone.} (P9)
\end{quotation}

These observations demonstrate that online communities, traditional media, and professional narratives, while differing in interactivity, comprehensiveness, and authority, each serve as essential hermeneutical resources that help individuals interpret their experiences and construct their identities. In some cases, however, participants encounter more immediate challenges to their beliefs, requiring short-term strategies of response, which we will discuss in the following.

\subsubsection{Finding support for own opinions in social environments}

When encountering cognitively dissonant content, the most common reaction among participants (reported by 8 out of 10) was to check comment sections or similar spaces to see if others had posted rebuttals to the original content and offered support for the views they shared. This strategy often provided emotional relief and a reassuring sense of ``not being alone'' in their response to the content. As P5 described, \textit{``It’s kind of like—I feel as if they’ve spoken on my behalf, said the things I wanted to say in the comments section. So in that sense, it feels a little better.''
}
% \begin{quotation}
% "If I see someone expressing the criticisms that I wanted to voice, I feel like there is someone who thinks the same way as I do, and that makes me feel more comfortable." (P6)
% \end{quotation}

% Beyond personal emotions, participants also valued moments when comment sections reflected broader social progress toward inclusivity, provided the discussions remained civil and focused on ideas rather than personal attacks. As P9 described, \textit{``If I holda  belief on something, and I see that other people are also supporting that idea, then I feel, Okay, the society is making progress; we’re on the right track now.' ''}

However, this strategy does not always work as expected. Participants often turned to comment sections seeking comfort, only to be met with its absence or the opposite. In many cases, this resulted in unresolved negative emotions and even greater distress. Recalling an incident when a prominent Taiwanese online personality was cyberbullied and forced to come out, P2 described feeling shocked and saddened to see the comment section flooded with homophobic remarks, which impacted his confidence towards the public acceptance of his identity:

\begin{quotation}
\textit{I'm already heartbroken from the incident itself, but what shocked me was that the people in the comments weren’t supporting him... I kept thinking, `How could people be like this? I guess there really still is such a large group of people who think that way.'}  (P2)
\end{quotation}

One reason for the lack of supportive speech lies in the limited visibility of affirming narratives within mainstream discourse on social platforms. Asexuality, for instance, remains relatively obscure in the Taiwanese online public sphere compared to topics like same-sex marriage and homosexuality. Yet triggers of cognitive dissonance for asexual individuals are pervasive, as mainstream narratives on social media often assume and celebrate romance and sexuality as ``default'' human desires. Reflecting this, P8, one of the few participants who did not view comment sections as a source of emotional relief, described the near impossibility for asexual people to find supportive narratives outside of their own communities:

\begin{quotation}

   \textit{When I come across a really hurtful comment, it's not like I can keep scrolling to see if I can find something more supportive. In reality, there just isn’t ... you can hardly find that kind of supportive comment outside of our own comfort zone. }(P8)
    
\end{quotation}

User-generated rebuttals in easily reachable places like comment sections offered LGBTQ+ individuals an important, though inconsistent, source of hermeneutical support within an unpredictable social environment. Despite their unreliability, such contributions illustrate a socially embedded means of providing hermeneutical resources to those in immediate need, suggesting a technological gap for intervention.

\subsubsection{Reaching out for resolution in own social spheres}

Participants sometimes search more proactively for a resolution by reaching out to their own social spheres. This often involves capturing the original content by screenshots or re-posting, and providing some personal remarks that critique the content or describe how they feel about it. This kind of reaching out can occur in public space, such as sharing an ``absurd'' hate speech to the individual's personal Instagram stories. 

More often, participants found deeper support and acknowledgment through private channels, such as direct message chats with close friends, secret or low-visibility social accounts, or conversations with intimate partners. Developing such strategies relied heavily on trust, disclosure, and the gradual accumulation of shared understanding of one’s life experiences and identities—an effortful process that required time and mutual sensitivity. As P8 explained, emotional support for them usually came through private chats with close friends: 

\begin{quote}
\textit{Because these are long-term relationships, I continuously share my life experiences and perspectives with them, and over time, they become more knowledgeable about it. At first, they were not very sensitive to these issues, so I had to do a lot of explaining... But gradually, as they listened, they developed this sensitivity, and now when I talk about hurtful things I see in the community, we can complain about it together, and stand on the same front.} (P8)
\end{quote}

% This mutual understanding of life experiences was described by many participants as the foundation for deeper exchanges of intimate feelings, particularly when discussing trauma or identity issues. As P5 explained, close friends provide a trusted space for disclosure because \textit{“we can talk about almost anything, because we know each other so well. We also know which topics are too sensitive to touch, such as certain traumas.”} By contrast, conversations with strangers lack this shared grounding, and thus \textit{“can easily overlook someone’s life experiences and cause harm.”}

Even with these considerations, this strategy sometimes produced unsatisfactory results. An unexpected lack of supportive feedback often led individuals to scrutinize the validity of their own opinions or the way they expressed them.

\begin{quotation}
\textit{When no one likes my post or story, I keep re-reading it and wonder what went wrong, was it my wording, or the way I expressed my thoughts?... Because I assumed some particular people would reply, but if they just read it without even liking it, I start to think the problem must be with me.} (P7)
\end{quotation}

Taken together, these accounts highlight both the potential and the fragility of reaching out as a strategy for cognitive and emotional resolution. On the one hand, strong ties and long-term mutual understanding with close friends can provide participants with a safe relational space for validation, solidarity, and even collective critique of hostile narratives. On the other hand, the very reliance on these strategies exposes participants to the risks of distress when the anticipated support fails to materialize. This contributes to a desire for participants to be intelligible with their narratives, in order to feel more entitled to express their lived experiences. 
% Nonetheless, the wide adoption and intuitiveness of this strategy also suggest opportunities for technological intervention.

\subsubsection{Fighting back in a discourse... or not?}

Encountering hostile or inconsiderate narratives on social media sometimes evokes the desire to respond directly. 
% The impulse to “fight back” is often tied to a sense of social responsibility to raise the visibility of their beliefs, or that silence would otherwise allow harmful discourses to persist unchallenged. 
Yet, participants emphasize that the decision to engage is not taken lightly, as it requires balancing emotional labor, the likelihood of meaningful dialogue, and the potential risks of further exposure to hostility.

% Others expressed a wish to use their responses as an opportunity for “micro-education,” even if the immediate effect might be limited. P10, for instance, often reposted hostile content to a private account, adding her own commentary to spark reflection among close friends:

% \begin{quotation}
% “I would usually share my thoughts along with the post—like pointing out the flaws in the logic, or using it as an opportunity to educate my friends. I don’t assume all of them are gender-conscious, but I still have some hope that by influencing people close to me, I can make small changes, even if I can’t change society at large.” (P10)
% \end{quotation}

For participants who do choose to engage, the quality and credibility of their responses are a central concern. Rather than posting emotionally charged reactions, some emphasized drafting carefully worded replies, fact-checking, and sourcing credible references before publishing. As P9 described:

\begin{quotation}
\textit{I usually draft my response, check if there are any problems, and verify the facts before posting. I want to convey accurate information, not just emotional outbursts. I don’t want people to think I’m untrustworthy or uninformed—because I really hate looking foolish.} (P9)
\end{quotation}

Some participants described motivation as a commitment to protecting their community outside of their comfort zone. P6, for example, notes that \textit{``why I respond is to pass on the knowledge to people within our own circles who share similar views. If we confront those misogynistic people directly, they won’t listen, because they’re too confident in their own rhetoric. So rather than consuming our energy on them, it’s better to focus on strengthening our own side.''}
% Despite these intentions, participants highlighted struggles in practicing this form of resistance. A recurring challenge was the lack of genuine space for dialogue. Many felt that those who posted or supported hostile content were not truly interested in discussion, but rather in provoking attention or reinforcing their own views. 
% P7 recalled being targeted when responding to gender-related debates:
% \begin{quotation}
% “When I responded, some people came at me, but they weren’t there for real discussion. Their views were already fixed, so there was hardly any room for dialogue. That just left me feeling powerless.” (P7)
% \end{quotation}
Participants repeatedly pointed to the burden of hermeneutical labor in online arguments as a gender minority. P8, reflecting on debates around asexuality, noted that engaging in prolonged ``comment wars'' often meant offering free education to an unwilling audience, while exposing oneself to more hostility:

\begin{quotation}
\textit{Arguing online just consumes your hermeneutical labor, and the other side won’t listen. It’s like you’re giving them free lessons, but they don’t care. Instead, it only exposes you to more hate speech. So in the end, I’d rather not bother.} (P8)
\end{quotation}

% Others further stressed that the form of online discourse itself often undermines the possibility of constructive engagement. P4 argued that overly academic or “bookish” rebuttals alienate general audiences, while confrontations in hostile spaces can quickly overwhelm the responder:

% \begin{quotation}
% “If you post a rebuttal in a page that holds totally different values, your comment just gets buried. And even when it’s people on the same side, if their posts are too bookish—full of academic jargon—it doesn’t work either. Online discussions need to be more plain and accessible. Otherwise, it’s not effective communication.” (P4)
% \end{quotation}

Taken together, these reflections illustrate the practice of fighting back as an important application of hermeneutical resources and a prominent site of hermeneutical labor. 
While some participants frame their responses as a form of advocacy, education, or community strengthening, the decision and benefit of responding is contingent on the quality and value of the response, as well as the perceived impact of one’s engagement. 
This points to an avenue of technology intervention where the effort to influence public discourse can be more efficiently directed and facilitated through the application of relevant hermeneutical resources.

% While these strategies highlight resilience and creativity, they also reveal fragility: supportive narratives can be absent, emotional labor is often unequally distributed, and the burden of responding to hostile discourse can become unsustainable. 
% These findings underscore the urgent need for technological interventions that can scaffold identity exploration, provide access to relevant hermeneutical resources, foster meaningful and sensitive discussions, and alleviate the cognitive and emotional demands of responding to dissonant or hostile content. Building on these insights, we next articulate four design goals to guide the development of an LLM-powered agent that addresses these challenges.

\section{Design Workshops \& Prototype}
Our formative study highlighted how Taiwanese LGBTQ+ individuals navigate social media through fragile yet creative strategies to access and apply hermeneutical resources. While these practices demonstrated resilience and self-determination, they also revealed limitations: supportive narratives are often absent, hermeneutical labor is unequally distributed, and cognitive burdens remain high. Through these findings, we identified a state of hermeneutical injustice in Taiwanese LGBTQ+ people's social media experience, {\color{blue}where LGBTQ+ user's ability to \textit{access} and \textit{apply} the hermeneutical resources to interpret or navigate their own social experience was impaired by social media designs in algorithm and interface. We also identified \textit{hermeneutical autonomy} as a core value within Taiwanese LGBTQ+ community that is currently poorly supported by platform mechanisms.} With this understanding, we next turned to a series of design workshops that translate {\color{blue}this value} into concrete design goals for a supportive system. We introduce the design process that leads to the construction of \textit{Queerbot}, an LLM-powered social support chatbot that aims to scaffold identity exploration, provide access to hermeneutical resources, and alleviate the cognitive and emotional demands of responding to hostility.

\subsection{Design workshops}
To make sure our design is aligned with the actual users, we held 3 workshops and invited users to participate in the design process.
Specifically, through the workshops, we aimed to (1) validate the scenarios in which intervention can be of help, (2) align the design goals of the prototype with the users, and (3) inspire solutions in the design.

\subsubsection{Participants}
8 participants who identified themselves as LGBTQ+ members took part in the design workshops. 2 of them (P8 \& P10) participated in the formative study. Their demographic information is shown in Table \ref{tab:dworkshop-demo}.

\begin{table*}[t]

\caption{Demographic information of the design workshops participants}
  \label{tab:dworkshop-demo}
  \Description{Table organized by three session IDs (D1, D2, D3) listing demographic details for eight participants. Columns include Participant ID, Age, Occupation, Gender, and Sexual Orientation. Session D1 includes one participant (P11). Session D2 includes three participants (P8, P12, P13), all students aged 22–24 with Non-binary or Agender identities. Session D3 includes four participants (P10, P14, P15, P16) aged 21–26, including students, a Designer, and a Project Manager. Sexual orientations across the groups include Pansexual, Bisexual, Asexual, Gynesexual, and Aromantic. Superscripts a and b mark specific terms in the Sexual Orientation column.}

  \begin{minipage}{\textwidth}
    \centering
    \begin{tabular}{ccclll}
      \toprule
      \textbf{Session ID} & \textbf{Part. ID} & \textbf{Age} & \multicolumn{1}{c}{\textbf{Occupation}} & \multicolumn{1}{c}{\textbf{Gender}} & \multicolumn{1}{c}{\textbf{Sexual Orientation}} \\ 
      \midrule
      \textbf{D1} & P11 & 21 & Studio Assistant & Non-binary & Aromantic\footnote{Describing a person who does not experience romantic attraction \cite{GLAAD-2022}.}, Pansexual \\ 
      \midrule
      \multirow{3}{*}{\textbf{D2}} 
                  & P8  & 24 & Student & Non-binary & Asexual \\
                  & P12 & 23 & Student & Non-binary & Gynesexual\footnote{Describing a person who is primarily sexually, aesthetically, and/or romantically attracted to femininity \cite{GLAAD-2022}.} \\
                  & P13 & 22 & Student & Agender    & Pansexual \\ 
      \midrule
      \multirow{4}{*}{\textbf{D3}} 
                  & P10 & 23 & Student & Cis Female & Pansexual \\
                  & P14 & 21 & Student & Non-binary & Pansexual \\
                  & P15 & 23 & Designer & Cis Female & Bisexual \\
                  & P16 & 26 & Project Manager & Cis Female & Bisexual \\ 
      \bottomrule
    \end{tabular}
  \end{minipage}

\end{table*}

\subsubsection{Materials}

Based on the findings of the formative study, we constructed 10 scenarios of common struggles related to hermeneutical injustice experienced by LGBTQ+ people, serving as the motivation for the design workshop. Sample scenarios include: \textit{unable to find responses that support my stances in comment sections}, \textit{struggling to interpret the discomfort or stress experienced on social media}, and \textit{feeling exhausted from repeatedly explaining my identity}. The complete list of scenarios can be found in Appendix \#2.

\subsubsection{Procedure}
All authors participated in the design of the workshop procedure, and the first author (FA) held the workshops across 1 week in July 2025 to ensure procedural consistency. The workshops took place offline in a meeting room at the university. Each workshop lasts approximately 100 minutes, and each participants were compensated with NT\$300 (about 10 USD) for their participation. {\color{blue}Participants were assigned to workshop sessions based on schedule availability.}

\begin{figure*}[t]
    \centering
    \begin{subfigure}[b]{0.40\textwidth}
        \centering
        \includegraphics[
            width=\textwidth,
            height=0.40\textheight,
            keepaspectratio
        ]{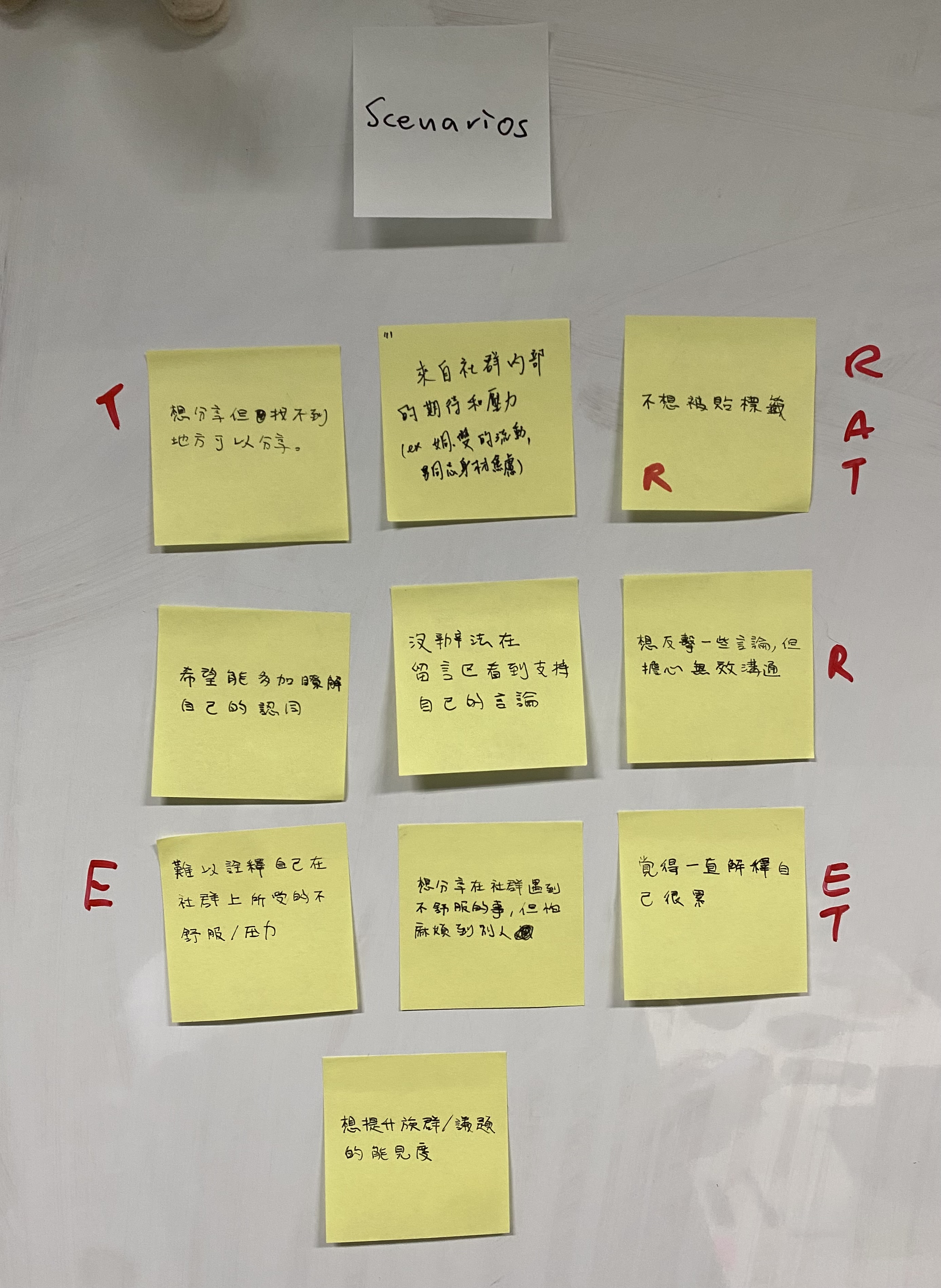}
        \label{fig:designWorkshopBoard1}
        \caption{Scenarios presented as design motivations.}
        \Description{Photo of a whiteboard containing ten yellow sticky notes with handwritten Chinese text, arranged under a top label reading "Scenarios," each referring to the scenarios elaborated in Appendix #3.}
    \end{subfigure}
    \hspace{0.04\textwidth} % controlled center spacing
    \begin{subfigure}[b]{0.40\textwidth}
        \centering
        \includegraphics[
            width=\textwidth,
            height=0.40\textheight,
            keepaspectratio
        ]{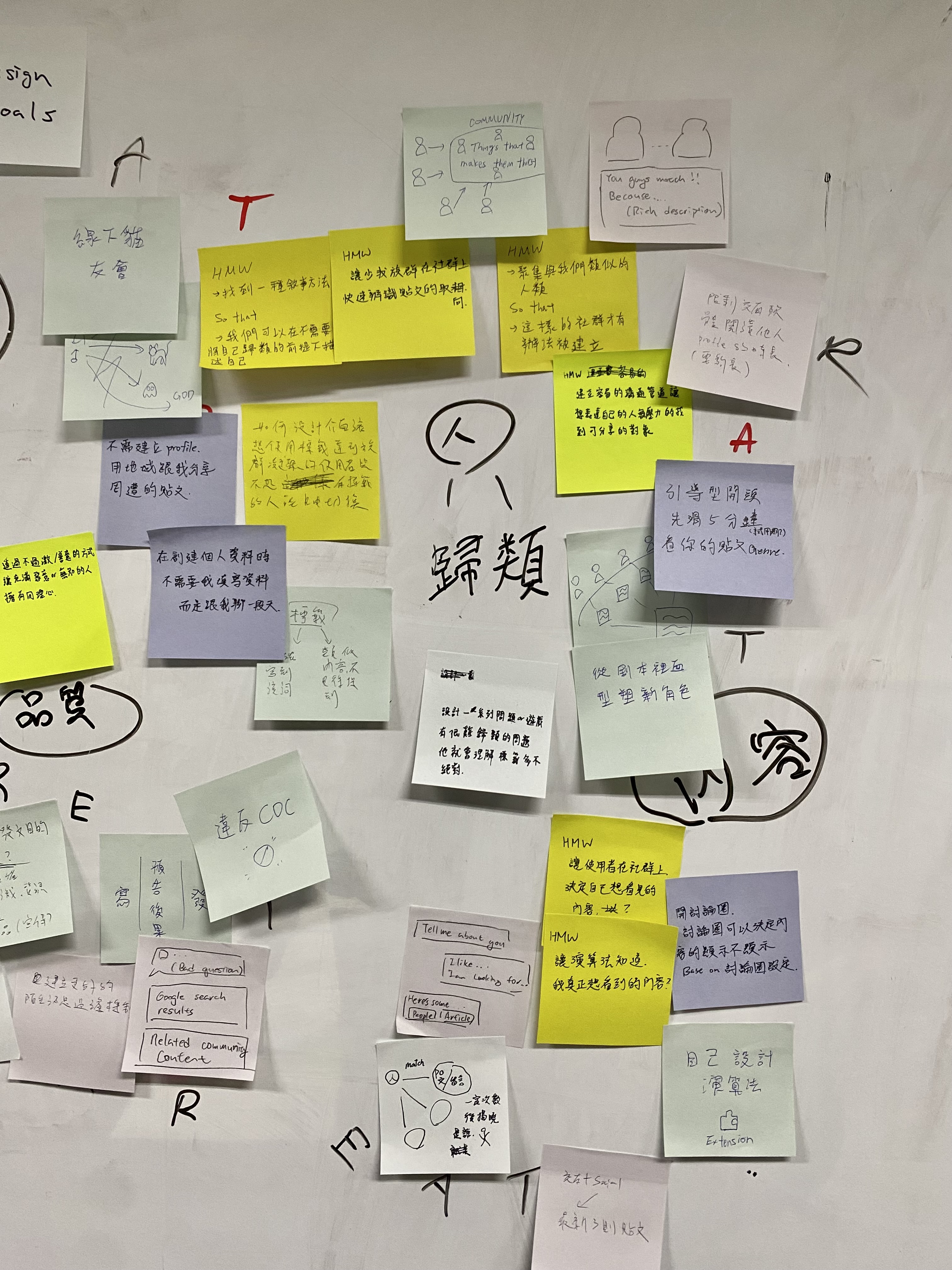}
        \label{fig:designWorkshopBoard2}
        \caption{"How might we"s with associated design ideas.}
        \Description{Photo of a whiteboard brainstorming session labeled "(b) 'How might we's with associated design ideas." The board is covered with yellow, purple, and light green sticky notes clustered around a central sketch of a stick figure labeled with the Chinese character for "Human" and the text "Categorization". Yellow notes primarily contain text labeled "HMW" (How Might We), while purple and light green notes feature a mix of text and interface wireframe sketches. Hand-drawn arrows and icons visually connect the various clusters and elements across the board.}
        
    \end{subfigure}

    \caption{Sections of workshop whiteboard captured from D3.}
    \label{fig:designWorkshopBoard}
\end{figure*}

%Note that, due to last-minute reschedule by other participants, D1 ended up hosting only one participant.

Each workshop started with approximately 5 minutes of check-in time, where the FA welcomed the participants and conducted small talks on the topic of the participants' hidden superpower to get acquainted with each other.

Next, the FA used approximately 10 minutes to introduce the results from the formative interview study to contextualize the study and familiarize the participants with the observed problem. Corresponding to the first goal of validating scenarios, the FA also encouraged the participants to raise any questions or personal experiences about the scenarios.

Next, corresponding to the second goal of aligning design goals, the FA asked the participants to write down sentences that started with \textit{``How might we''} that address the given scenarios on post-it notes and share them with each other. They were then instructed to sort the notes based on common themes and then collectively select 3-4 design goals.

After the design goals have been selected and agreed upon, the participants would spend about 10 minutes sketching out ideas on paper that try to solve the problems and sharing them with the team, jotting down any feedback they received and deemed important. This solution brainstorming session corresponded to the third goal of inspiring solutions and was iterated for three rounds, taking up about 30 minutes in total. In this session, the FA had prepared some pre-made user interface components (e.g., chat boxes and social media posts) for the participants to use in their sketching.

% Next, the FA presented a demo video of Queerbot in an early development stage. Rather than treating this as a finalized design, the demo was framed as the background and basis for the participants’ subsequent brainstorming. The FA invited participants to consider: if this prototype were to be further developed, what features, functions, or interactions would make the agent more useful in supporting their needs? This encouraged participants to situate their ideas within a design direction while still leaving room for critical reflection and creative expansion.

The workshops were concluded with a 10-minute brainstorming session, where the participants ideated individually to sketch a prototype that brings together their favorite ideas {\color{blue}from the brainstorming sessions}. 
% \todo{Timmy: please add some photos here.}
%either from the ideas that were generated from the brainstorming sessions or the hi-fi prototypes we presented.

\subsubsection{Data analysis}
After each of the design workshops, the participants and the FA together organized the generated ideas into clusters and themes based on the underlying desire they addressed. Based on these results, the FA did the first round of coding on these ideas and engaged in active discussion with the rest of the authors for agreement. 17 codes were generated in the process, 4 dimensions emerged, which we collectively decided were most relevant, and we present them in the following.

\subsubsection{Findings}
Among the themes that emerged from our analysis, {\color{blue}four dimensions were particularly informative for the prototype design: (1) \textbf{Self}, (2) \textbf{People}, (3) \textbf{Information}, and (4) \textbf{Environment}. Since the themes of People and Information show conceptual similarity, we present them within the same section. We also found that the majority of ideas in the themes of People, Information, and Environment can also be situated within the theme of Self, which together informed three jointly constructed design directions and subsequent design goals (see Fig. \ref{fig:designWorkshopResults}). Description of themes and example ideas of each theme can be found in Appendix \#4.} We elaborate on these dimensions in the following:

\begin{figure*}[t!]
    \centering
    \includegraphics[width=0.95\linewidth]{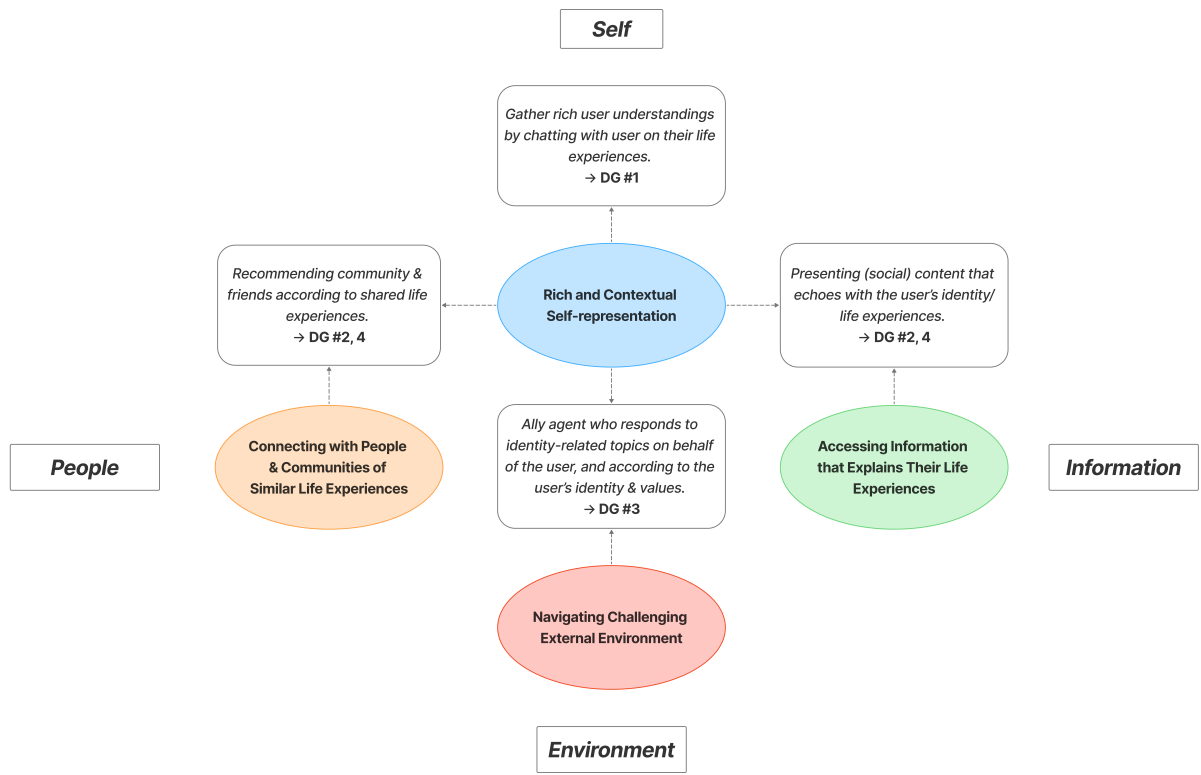}
    \caption{Identified themes (rectangles), latent desires (ovals), and design directions along with the design goals they inform (rounded rectangles) from the findings of design workshops.}
    \Description{Diagram mapping design themes, desires, and directions around a central blue node labeled "Rich and contextual self-representation." Four axes labeled Self, People, Information, and Environment surround this center. At the top under "Self," a node reads "Gather rich user understandings by chatting with user on their life experiences (DG #1)." To the left under "People," an orange node regarding connecting with similar communities points to a node reading "Recommending community \& friends according to shared life experiences (DG #2, 4)." To the right under "Information," a green node regarding accessing explanatory information points to a node reading "Presenting (social) content that echoes with the user's identity/life experiences (DG #2, 4)." At the bottom under "Environment," a pink node regarding navigating challenging environments points to a node describing an "Ally agent who responds to identity-related topics on behalf of the user, and according to the user's identity \& values (DG #3)." All four directions connect inward to the central blue node.}
    \label{fig:designWorkshopResults}
\end{figure*}

\textbf{The Self.} Participants articulated that an agent should not treat identity as a fixed set of categories but as something fluid, contextual, and relational. Their visions for the system can be grouped into two overarching dimensions: (1) fluid and contextual self-representation, where users dynamically express, refine, and explore identities in interaction with the agent and others; and (2) mediated identity practices and stewardship, where the agent facilitates safer, more meaningful engagements with others while also relieving hermeneutical labor.

% \paragraph{Fluid and contextual self-representation.}
Participants expressed strong interest in mechanisms that allow identity to be represented in ways that are flexible, layered, and situational. Rather than filling out rigid forms, they preferred conversational interfaces that let them ``tell'' the system about themselves in natural language, combining labels with free-text self-descriptions and inner qualities. They also emphasized transparency and control, suggesting features such as editable ``what I know about you'' views of user knowledge. Beyond profiles, participants wanted the agent to adapt to evolving social experiences, for instance, by suggesting relevant groups or communities after a user shares personal narratives, or by highlighting resonances with others (\textit{``you guys match!! because…''}). These practices underscore participants' desire for identity to be something consciously negotiated, not implicitly inferred.
% Playful experimentation was also important: some envisioned role-play features where they could adopt alternative identities—such as other genders, fictional characters, or even non-human roles—within structured games or low-stakes scenarios. 

% \paragraph{Mediated identity practices and stewardship.} 

% The second dimension concerned the ways the agent might mediate identity-related interactions. 
% To reduce conflict and support emotional safety, participants proposed mechanisms such as previews of the “status” of comment sections (e.g., \textit{“flaming war”} vs. \textit{“peaceful zone”}), or pre-defined discussion spaces categorized by levels of conflict. 
% They also suggested mechanisms to encourage engagement with others’ lived experiences, such as surfacing long-form self-expression about identity struggles. 
% At a broader level, participants extended the agent’s role to identity stewardship, imagining it as a digital twin that could speak on their behalf, relieving the burden of repetitive explanation and education about their identity. This situates the agent as both an immediate support and a custodian of their identity work.

\textbf{The People \& Information.}
Participants emphasized that meaningful support requires attending to both the community users can connect with and to what informational resources they can access for interpretation. The two often interweaving dimensions highlight the need for (1) fostering belonging through community connections and (2) scaffolding understanding through contextualized information.

\paragraph{Community resources.}
Participants stressed the importance of accessing communities that help them understand themselves with similar life experiences, based on nuanced identity understanding. Instead of relying solely on static identity categories, they envisioned systems that could interpret lived experiences and then recommend supportive communities, groups, or peers. Features that highlight ``others with similar experiences'' or suggest identity terms based on shared struggles were seen as crucial to helping users locate solidarity. By grounding recommendations in the richness of personal experiences, participants hoped the agent could enhance a sense of belonging and reduce the isolation that many LGBTQ+ individuals face online.

\paragraph{Informational resources.}
Participants also underscored the importance of informational support as part of hermeneutical scaffolding. They wanted access to resources that are contextualized, personalized, and responsive to lived experience. For instance, after describing a social encounter, users expected the agent to contextualize their experience in a rational manner and recommend relevant terms, groups, or theoretical frameworks that could help them interpret the situation. They also imagined more conversational discovery mechanisms, such as being able to answer prompts like \textit{``Tell me about you''}, with the system surfacing relevant resources and articles in response. These expectations highlight the need for systems that act not just as neutral conduits of information, but as interpreters that can support autonomous access and application of hermeneutical resources that bridge between personal experiences and broader narratives.

\textbf{The Environment.}
Participants envisioned the agent not only as a source of information but also as an ally for navigating difficult or hostile environments. In these contexts, the agent was expected to help users filter poor-quality input, confront challenging discussions, and even carry some of the communicative burden on their behalf. 
% Their suggestions can be grouped into two themes: (1) allyship and supportive presence in challenging environments, and (2) quality control and guidance in navigating problematic discourse.

% \paragraph{Allyship and supportive presence.}
Some participants indicated the need for the agent to act as a trusted ally when facing repetitive or hostile communication. This could involve an “ally bot” that helps handle repetitive exchanges, or one that explicitly signals readiness to accompany the user in difficult spaces (e.g., “\textit{Let’s fight! Ally bot is ready!”} after a user chooses to engage in a challenging social environment). Other participants proposed designs that help users navigate problematic or unproductive conversations by improving the quality of discourse and providing educational resources instead. One suggestion was that when an inconsiderate or ``bad'' question is detected, the system could redirect the attention of the author by automatically displaying relevant Google search results and community resources, thereby reducing the burden on users to provide explanations. 

{\color{blue}At a more existential level, participants also imagined the agent as a digital twin that could represent their perspectives and answer identity-related questions on their behalf. One participant, P14, imagined having this kind of proxy to act as a “digital gravestone” after its death, serving as an active memorial that portrays and reflects the personality it once was.} These visions highlight the desire for the agent to extend beyond information delivery and actively embody identity in the face of social adversity.
% Similarly, participants imagined a rating system for questions, where users could evaluate prompts with five stars, and low ratings would trigger reflective follow-ups (e.g., \textit{``Should Google first,'' ``Too generalized,'' or ``Other''}). These mechanisms would not only regulate the environment but also encourage learning, nudging participants toward more constructive contributions while reducing unnecessary hermeneutical labor.

\begin{figure*}[h]
    \centering
    \includegraphics[width=0.88\linewidth]{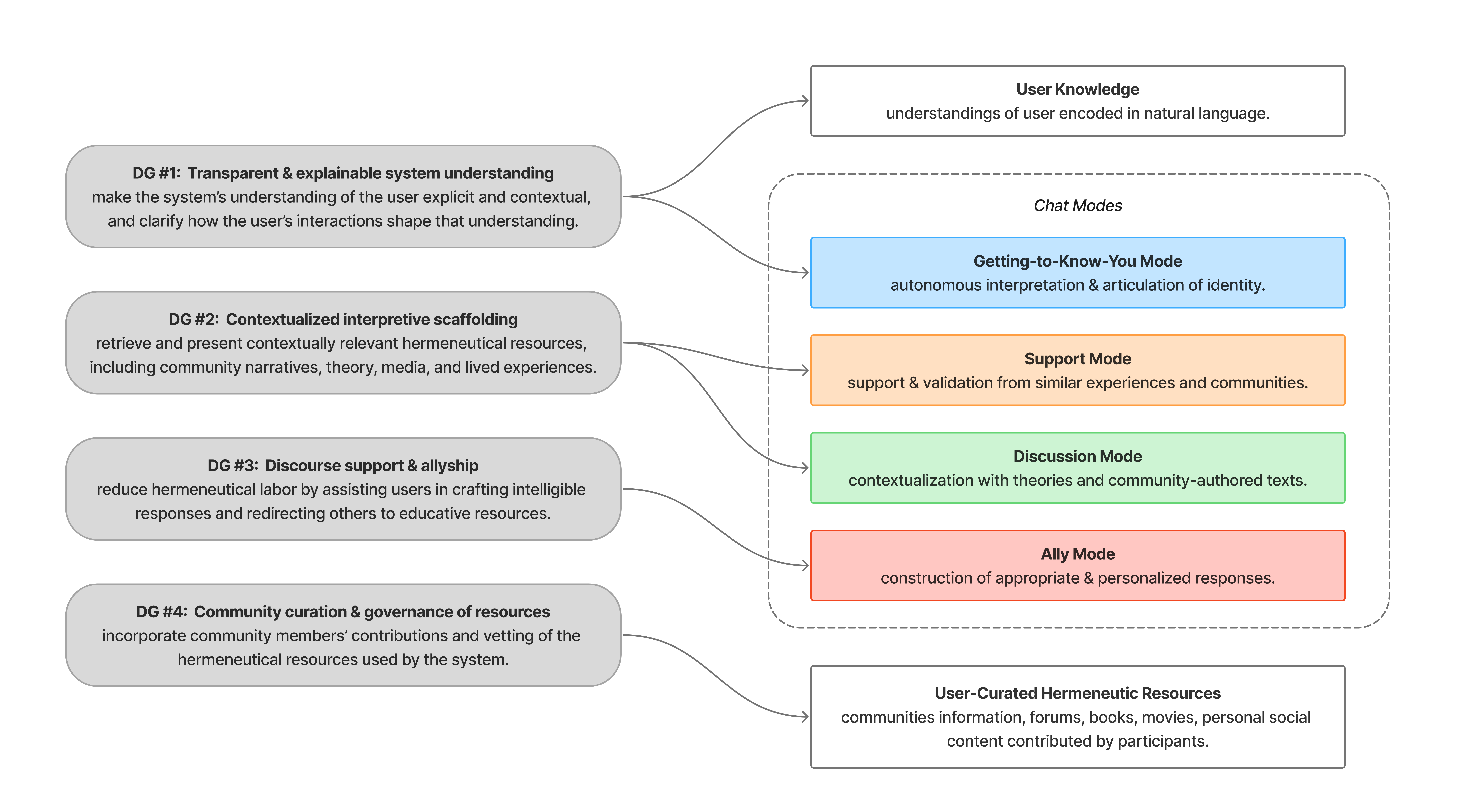}
    \caption{Mapping of design goals (rounded rectangles) to the feature \& functionalities of Queerbot (rectangles)}
    \Description{Diagram mapping four Design Goals (DG) on the left to corresponding system features on the right. DG #1 (Transparent & explainable system understanding) connects to "User Knowledge" and the blue "Getting-to-Know-You Mode." DG #2 (Contextualized interpretive scaffolding) connects to the orange "Support Mode" and green "Discussion Mode." DG #3 (Discourse support \& allyship) connects to the red "Ally Mode." These four modes are grouped within a dashed box labeled "Chat Modes." Finally, DG #4 (Community curation & governance) connects to "User-Curated Hermeneutic Resources" at the bottom.}
    \label{fig:dgToFeature}
\end{figure*}

\subsection{Design goals}
{\color{blue}Based on the results of the design workshop, here we list the design goals (DGs) our prototype should achieve, as well as the findings from design workshops they address. DGs \#1-3 articulated the core requirements for an agent that meaningfully addresses the challenges faced by LGBTQ+ individuals on social media in terms of access and application of hermeneutical resources, while DG \#4 reflected participants' underlying desire for community control over the epistemic infrastructure that influences them.}

\begin{itemize}
    \item \textbf{DG \#1: Transparent \& explainable system understanding.} Make the system’s understanding of the user explicit and contextual, and clarify how the user's interactions shape that understanding. {\color{blue}This primarily addresses the value of fluid and autonomous self-representation surfaced in the theme of \textbf{the self} in our design workshops, and serves as the foundation for subsequent interactions.}
    
    \item \textbf{DG \#2: Contextualized interpretive scaffolding.} Retrieve and present contextually relevant hermeneutical resources, including community narratives, theory, media, and lived experiences, with particular attention to underrepresented identities. {\color{blue}This supports the theme of \textbf{the people \& information} from the design workshops, where participants imagined gaining meaningful support from both community and informational resources.}
    
    \item \textbf{DG \#3: Discourse support \& allyship.} Reduce hermeneutical labor by assisting users in crafting intelligible responses and redirecting others to educative resources. {\color{blue}This goal responds to the struggles and desires participants described when engaging with difficult or hostile environments, as articulated in the theme of \textbf{the environment} in our design workshop findings.}
    
    \item \textbf{DG \#4: Community curation \& governance of resources.} Incorporate community members’ contributions and vetting of the hermeneutical resources used by the system. {\color{blue}This goal is supported by our empirical observation that participants wished for community-based resource support in the theme of \textbf{the people \& information} in our design workshops, and is further informed by prior calls for community-driven design of epistemic infrastructures that align system affordances with the values and practices of traditionally marginalized groups \cite{KayEtAl-2025, WanEtAl-2023}.}
\end{itemize}

% They emphasize the need for flexibility in identity exploration, access to diverse and credible hermeneutical resources, sensitivity and inclusivity in dialogue, and assistance in managing the labor of responding to hostile or dissonant content.  To operationalize these goals, the system must be underpinned by a technical architecture capable of integrating personalized knowledge management with robust mechanisms for retrieving and contextualizing relevant resources. 
In the next section, we describe the prototype features, architecture, and data structures, and how they support these design goals.

\subsection{Features \& functionalities}

We designed and implemented Queerbot, a LLM-driven social support chatbot, to address the aforementioned design goals. Here we present the features and functionalities of Queerbot, along with the design goals it addresses.

\subsubsection{User Knowledge}

\begin{figure}[h]
    \centering
    \includegraphics[width=0.4\linewidth]{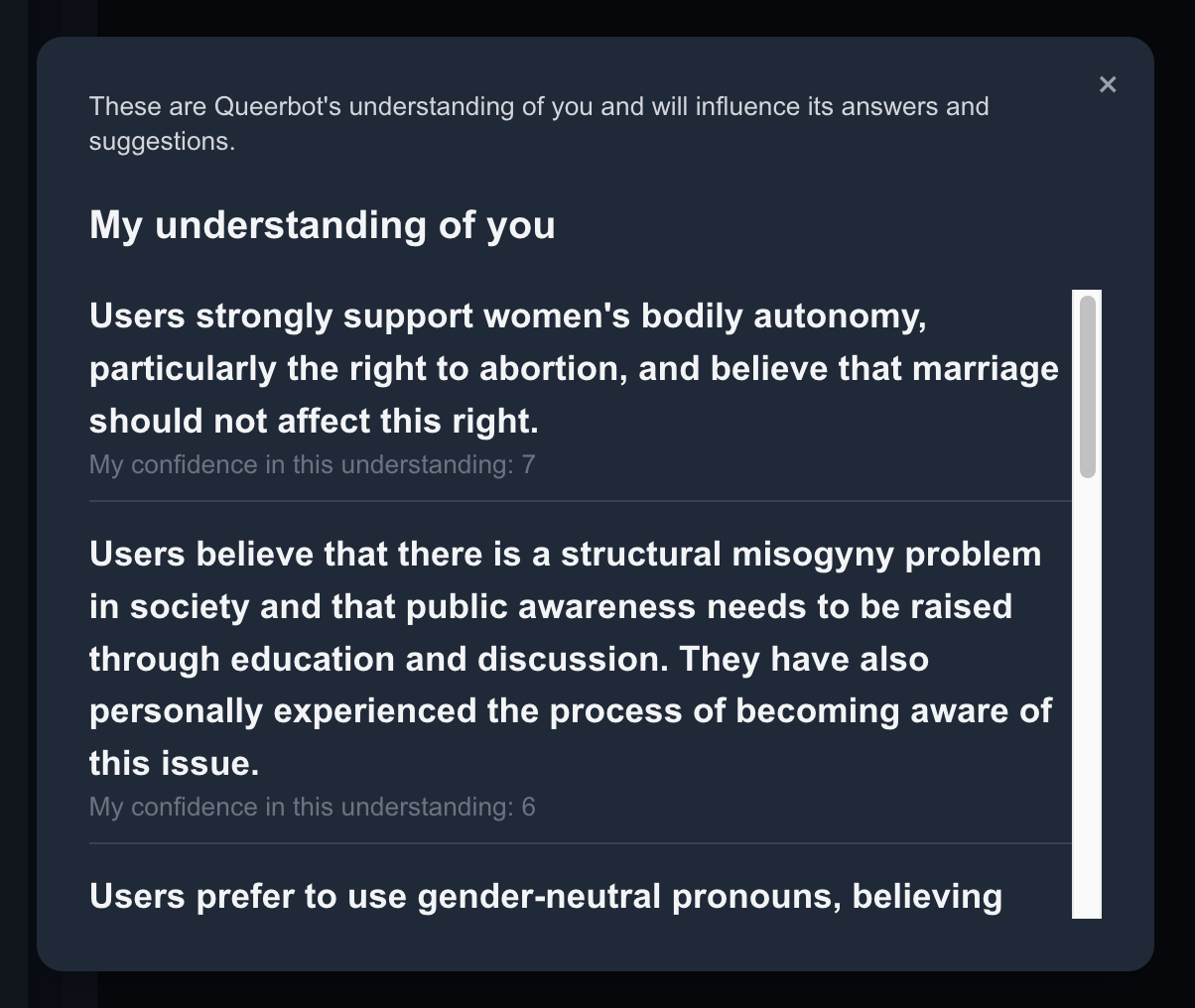}
    \caption{User knowledge displayed by Queerbot}
    \Description{Screenshot of a modal window titled "My understanding of you" listing the system's inferences about the user. A scrollable list displays natural language statements summarizing user beliefs, each accompanied by a numeric confidence score. Visible examples include statements regarding support for women's bodily autonomy (Confidence: 7), belief in structural misogyny and the need for public awareness (Confidence: 6), and a preference for gender-neutral pronouns.}
    \label{fig:userKnowledge}
\end{figure}

To enhance transparency and control of system understandings as required by DG \#1, the prototype provides a list of its understanding about the user (\textit{``What we know about you''}), as well as a post-conversation feedback that summarizes updates to the user’s knowledge profile. This allows participants to monitor how their persona is represented in the system and reflect on their evolving self-understanding.

\subsubsection{Chat Modes that Serve Different Needs}

\begin{figure*}[t!]
    \centering
    \includegraphics[width=0.9\linewidth]{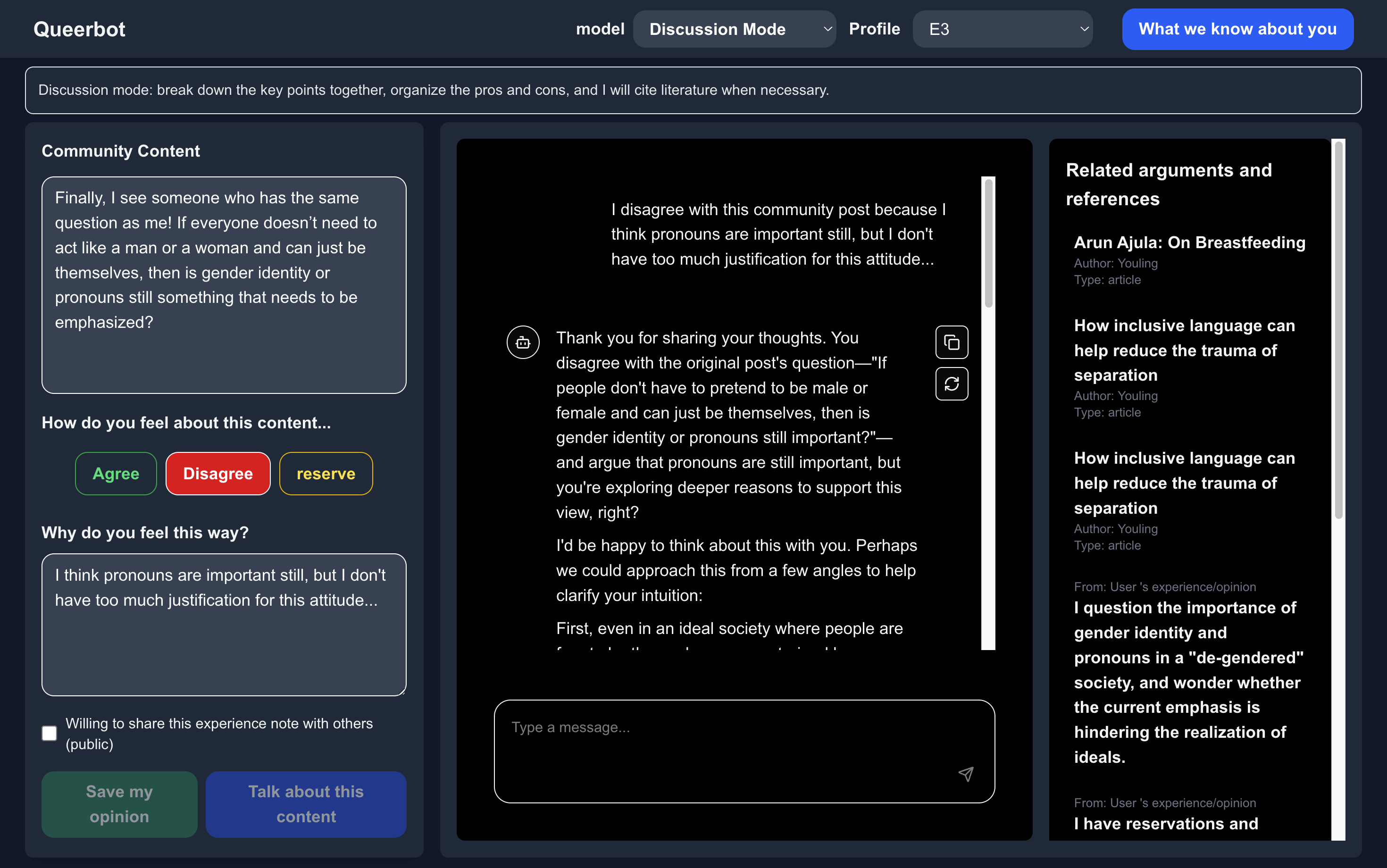}
    \caption{Queerbot in discussion mode}
    \Description{Screenshot of the Queerbot interface in "Discussion Mode" featuring a three-column layout. The left panel displays a "Community Content" post for the user to evaluate, containing buttons to Agree, Disagree, or Reserve, and a text field for inputting reasoning. The central panel is a chat interface showing an active conversation where the bot helps the user analyze their input. The right panel lists "Related arguments and references," displaying citations and snippets of relevant literature or user opinions.}
    \label{fig:discussionMode}
\end{figure*}

To support the core requirements dictated by DGs \#1-3, Queerbot was configured to act differently and base its response in different contexts in four distinct chat modes.

\begin{itemize}
    \item \textit{Getting-to-Know-You Mode}: In this mode, the chatbot proactively asks questions to fill knowledge gaps and extend its understanding of the user. This works in tandem with \textit{User Knowledge} in addressing DG \#1 by supporting autonomous interpretation of identity by prompting users to articulate their experiences, preferences, and stances.
    \item \textit{Support Mode}: When users encounter difficult or hurtful social content, the chatbot supports DG \#2 by focusing on acknowledgment and validation, drawing on similar experiences contributed by other users. It also recommends hermeneutical resources such as communities, books, websites, and movies that may help users feel less isolated and better understood.
    \item \textit{Discussion Mode}: For users who wish to process content more deeply, the chatbot supports DG \#2 by contextualizing sociopolitical issues with theories and community-authored texts. This mode helps users untangle complexity, understand different perspectives, and understand their own positions. Through the provision and application of more in-depth hermeneutical resources, the bot provides cognitive distance and critical lenses for reflection on identity-related issues.
    \item \textit{Ally Mode}: In encounters with inconsiderate or hostile social content, the chatbot addresses DG \#3 by assisting users in constructing responses aligned with their identity and lived experiences. The system presents multiple response options, supported by relevant hermeneutical resources, so that users can choose a strategy that balances emotional labor with advocacy goals.
\end{itemize}

Together, these modes scaffold both the access and application of hermeneutical resources, while giving users agency to decide how they want the chatbot to respond in different contexts.

\subsubsection{User-Curated Hermeneutical Resources}

Finally, the prototype addresses DG \#4 by incorporating hermeneutical resources directly contributed by participants. These include LGBTQ+ communities, online forums, books, movies, and personal social content that participants identified as meaningful in making sense of their experiences. By integrating community-grounded resources alongside academic and professional ones, the chatbot offers responses that feel authentic, situated, and sensitive to the lived realities of LGBTQ+ individuals.

\subsection{Architecture and data}

Queerbot is a chatbot designed around the RAG (Retrieval Augmented Generation) technique, a way to improve the quality of domain-specific responses by combining the retrieval of curated knowledge with the paraphrasing capabilities of LLMs \cite{FanEtAl-2024}. We also used a simplified version of the approach proposed by \cite{ZhaoEtAl-2024} to create, update, and manage user knowledge, in order to provide personalized experiences. This section describes the data flows supporting our chatbot.

\begin{figure*}[h]
    \centering
    \includegraphics[width=0.65\textwidth]{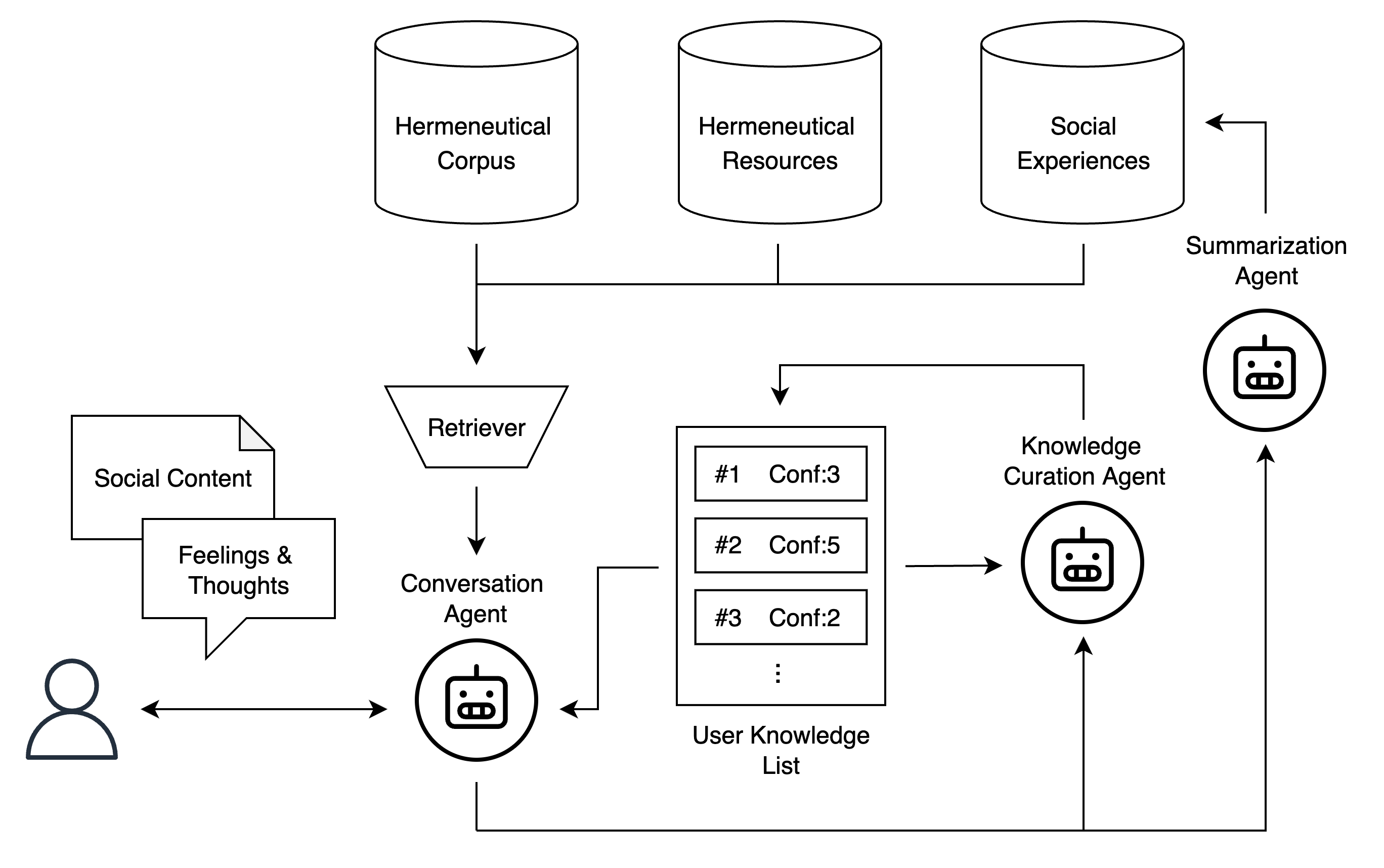}
    \caption{Data flow of Queerbot}
    \Description{Diagram illustrating the data flow between system components. Three databases at the top (Hermeneutical Corpus, Hermeneutical Resources, Social Experiences) feed into a "Retriever." The Retriever supplies a central "Conversation Agent," which interacts bidirectionally with a User on the left (exchanging "Social Content" and "Feelings & Thoughts"). The Conversation Agent updates a central "User Knowledge List" containing items with confidence scores. On the right, a "Knowledge Curation Agent" refines this list, while a "Summarization Agent" captures interaction data to update the "Social Experiences" database at the top.}
    \label{fig:dataFlow}
\end{figure*}

\subsubsection{Data stores}

Data stores provide the chatbot with personalization as well as hermeneutical resources, as shown in Figure \ref{fig:dataFlow}. Except for User knowledge, which was attached in the generation context as a complete list, all data listed here are indexed for retrieval, and only appear in the generation context when it's relevant to the particular prompt.

{\color{blue}

\begin{itemize}

\item \textit{User knowledge}: A collection of atomic descriptors about the user, capturing relevant aspects of their identity, preferences, opinions, and values. Each item is associated with a dynamic ``confidence'' score representing Queerbot’s current degree of belief in that piece of knowledge. All items are initialized at a score of 2. Confidence scores can be incremented or decremented through ongoing user–bot interactions, and any item whose score reaches 0 is removed from the knowledge store.

\item \textit{Social experiences}: Concise summaries of past interactions between the user and the bot. These entries may include user-provided social context, their reflections on the situation, and the ensuing discussion. With the user’s explicit consent, these summaries can be optionally shared with other users to support community learning.

\item \textit{Hermeneutical resources}: Curated excerpts of interpretive materials, such as websites, community spaces, films, and books, which offer conceptual, value, or identity-relevant grounding. These resources are available to all users.

\item \textit{Hermeneutical corpus}: Full-text versions of hermeneutical resources, including authoritative gender theories and articles authored by members of local LGBTQ+ communities. Like the resources above, these materials are shared across all users.

\end{itemize}

\subsubsection{Response generation.}
The LLM generates responses using three contextual inputs: the user profile (constructed from the User Knowledge store), the mode-specific resource context (retrieved hermeneutical resource collections), and the mode-specific prompts (behavioral scaffolds that guide the system’s tone and strategy).

\subsubsection{Data updates.}
After each interaction, a summarization agent produces a summary of the interaction. Based on the interaction, a knowledge curation agent updates the User Knowledge store by adding new items, revising existing ones, or adjust confidence scores to reflect the change of user attributes.

}

\subsection{System implementation}
The user interface and core logic were implemented with Next.js Web framework, with the help of the Langchain.js
package. Supabase was used to store our data and serve as vector stores for retrieval. We leveraged Gemini 2.5 Flash for
LLM inference and gemini-embedding-exp-03-07 text embedding, both via the Gemini API.

\section{Evaluation Workshops}
With the aim to evaluate the potential of our prototype in supporting hermeneutical autonomy and to elicit insights into related lived experiences of Taiwanese LGBTQ+ people on social media, we conducted a total of 5 evaluation workshops. Participants were invited to engage with the refined prototype and discuss the feelings and opinions evoked during the experience. Crucially, the discussion following these tasks prioritized reflections on participants' latent desires and struggles over system usability. The evaluation workshops were designed with three main goals: (1) to validate whether the design direction resonated with potential users, (2) to evaluate the prototype design and its features in supporting nuanced identity-related interactions, and (3) to uncover deeper desires and struggles embedded in their situated realities that may not have been fully captured in earlier design workshops.

\subsection{Method}

\subsubsection{Participants} We recruited 11 participants of the LGBTQ+ identities. Participants were recruited through community connections and a public post by the FA on Threads, which was re-posted by dozens of local LGBTQ+ community members and organizations. None of these participants overlapped with the earlier formative interviews or design workshops. We conducted five workshop sessions, each consisting of two to three participants. Each session lasted approximately 90–100 minutes, and all participants received NT\$300 (about 10 USD) compensation for their time and contributions. The demographic details of participants are presented in Table \ref{tab:eworkshop-demo}.

\begin{table*}[h]
  \caption{Demographic information of the evaluation workshops participants}
  \label{tab:eworkshop-demo}
  \Description{Table organized by five session IDs (E1–E5) listing demographics for eleven participants (P17–P27). Columns include Participant ID, Age, Occupation, Gender, and Sexual Orientation. Ages range from 19 to 31. Occupations are primarily Students, alongside a Public Servant, Gym Executive, Travel Agent, and Caregiver. Gender identities include Cis Female, Cis Male, Trans Female, Agender, Non-binary, Queer, and Transgender. Sexual orientations span Homosexual, Pansexual, Demisexual, Androsexual, Bisexual, Asexual, and Sapiosexual. Superscripts a through c mark specific terms in the gender and orientation columns.}
  
  \begin{minipage}{\textwidth}
    \centering
    \begin{tabular}{ccclll}
      \toprule
      \textbf{Session ID} & \textbf{Part. ID} & \textbf{Age} & \multicolumn{1}{c}{\textbf{Occupation}} & \multicolumn{1}{c}{\textbf{Gender}} & \multicolumn{1}{c}{\textbf{Sexual Orientation}} \\ 
      \midrule
      \multirow{2}{*}{\textbf{E1}} 
        & P17 & 20 & Student & Cis Female & Homosexual \\
        & P18 & 22 & Student & Agender\footnote{Describing a person who do not identify with any particular gender or has very little experience of a gender \cite{APAStyle-, CambridgeUniversityPress-n.d.}.} 
                                 & Pansexual, Demisexual\footnote{Describing a person who feels sexually attracted only within the context of a strong emotional connection with another person \cite{APAStyle-}.} \\ 
      \midrule
      \multirow{2}{*}{\textbf{E2}} 
        & P19 & 21 & Student & Non-binary & Androsexual \\
        & P20 & 21 & Student & Queer      & Bisexual \\ 
      \midrule
      \multirow{3}{*}{\textbf{E3}} 
        & P21 & 19 & Student & Trans Female & Androsexual \\
        & P22 & 25 & Public Servant & Queer & Bisexual \\
        & P23 & 30 & Gym Executive & Non-binary, Queer & Pansexual \\ 
      \midrule
      \multirow{2}{*}{\textbf{E4}} 
        & P24 & 27 & Student & Cis Female & Bisexual, Asexual, Sapiosexual\footnote{Describing a person who is sexually or romantically attracted to intelligent people \cite{CambridgeUniversityPress-n.d.a}.} \\
        & P25 & 31 & Student & Cis Male   & Homosexual \\ 
      \midrule
      \multirow{2}{*}{\textbf{E5}} 
        & P26 & 30 & Travel Agent & Cis Female & Homosexual \\
        & P27 & 25 & Caregiver    & Transgender, Non-binary & Homosexual \\ 
      \bottomrule
    \end{tabular}
  \end{minipage}
\end{table*}

\subsubsection{Materials} To ensure the scenarios and experiences are grounded in personal and community realities, we collected two types of workshop materials: \textit{personalized inputs} and \textit{real-world social content stimuli}.

\begin{itemize}
    \item \textit{Personalized inputs.} Prior to each session, we collected journal entries, past social content, and community resources from participants through dedicated LINE chatrooms. These materials were incorporated into the system to personalize the prototype experience.
    \item \textit{Real-world Social content stimuli.} The authors also curated a set of real-world social media posts that reflected discriminatory, stereotyped, or ambiguous gender-related content. These posts were pseudonymized to protect privacy and adapted into a lo-fi Figma prototype resembling a Threads-like social media interface. For tasks 2–4, participants were presented with 2–3 options of such content within this prototype, and copied the one they wished to discuss into the chatbot.
\end{itemize}

\subsubsection{Procedure}

All authors collaborated in designing the procedure. Each session began with a short warm-up activity (about 5 minutes), where participants engaged in small icebreaker games to build comfort with one another.

The facilitators then provided a brief introduction (about 5 minutes) to the purpose of the chatbot and the evaluation workshop. Participants were introduced to the four prototype modes and the tasks they would engage in. Worksheets containing design goals, scenario descriptions, screenshots, and reflection questions were distributed, and ground rules regarding respect, confidentiality, and openness were established.

The core of the workshop consisted of four evaluation tasks (about 100 minutes in total). In Task 1a (Knowledge Verification), participants reviewed a list of user knowledge items accumulated by the system based on their prior content. They marked whether statements were accurate, deleted incorrect ones, and added new information. In Task 1b (Getting-to-Know-You Mode), participants interacted with the chatbot in a conversational style, experiencing how the chatbot asked reflective questions to deepen its understanding of their identity.

For the remaining tasks, participants interacted with the chatbot by copying a material from the Figma prototype, and followed instructions from the chatbot. In Task 2 (Support Mode), they selected a discriminatory post and explored the supportive response and resources suggested by the chatbot. In Task 3 (Discussion Mode), they chose a complex or mildly inappropriate post and asked the chatbot to contextualize the content. In Task 4 (Ally Mode), they selected a challenging post and explored how the chatbot generated allyship-oriented responses, including different phrasing options. After each task, participants wrote down their reflections on sticky notes and posted them on a large A0 poster visible to the whole group. This setup allowed participants to see others’ responses and build upon them during the discussion.

Finally, the workshop concluded with a group reflection and debrief (10 minutes). The participants shared their impressions of each mode, discussed the strengths and limitations of the chatbot, and identified important scenarios that are not addressed by this prototype.

{\color{blue}\subsection{Data Analysis}

All workshop sessions were audio-recorded and transcribed with participants’ consent. We conducted an inductive thematic analysis \cite{BraunClarke-2006, TaylorBruckman-2024} on the transcripts. The analysis began with the first and second authors, who also served as workshop facilitators, independently reviewing the data to generate an initial set of codes. The two coders then engaged in a series of consensus meetings to compare their coding, resolve discrepancies through discussion, and cluster the codes into preliminary themes. These themes were subsequently refined and finalized through iterative discussion with the full research team. The complete codebook is provided in Appendix~\#5. The findings presented below reflect the recurring patterns and salient insights observed across participants.}

\subsection{Findings}

\subsubsection{Reflection, re-evaluation, and exploration of identity}

Participants repeatedly emphasized how Queerbot's curated user knowledge and responses prompted them to proactively reflect on their current self-understanding. For many, the system’s ability to synthesize scattered inputs into descriptive observations and tailored conversations functioned like a mirror, revealing underlying traits or orientations they had not explicitly articulated.
% Participants also described how the daily reflections they conducted before the workshop encouraged engagement with issues they might otherwise neglect. 

Labels were a recurring theme. Some acknowledged the social necessity of labeling: \textit{``to live in today’s society, I still need to give myself some labels''} (P20), and the potential of further exploration prompted by a short term. These categorical nouns, however, were often seen as inadequate for capturing rich, complex identity. Longer descriptive sentences offered by Queerbot were valued as a more nuanced alternative to labeling, allowing self-positioning and exploration without rigid confinement.

Participants also wanted the chatbot to more actively push their thinking with relevant questions. For example, by surfacing terms like “misogyny” that the participant had not explicitly mentioned when discussing an unpleasant social encounter, it offered a sharper lens for reflection: \textit{``It’s like giving me a topic or an issue to explore, which then allows me to extend and reflect more on my own stance''} (P19). 
% Finally, the temporal aspect mattered: meaningful changes in self-identity were seen as requiring longer-term observation.  

% These results revealed participants' desire to reclaim the autonomy in identity formation and exploration that social media algorithms had undermined, restoring space for self-definition through proper accessing of hermeneutical resources like terms and concepts. 

\subsubsection{Validation, actionable guidance, and not being alone}

Participants highlighted the importance of Queerbot in offering validation, reassurance, and a sense of \textit{not being alone} through interpretive resources. For example, when confronted with dismissive comments like \textit{``who cares about your gender?''}, the AI reframed these experiences by first affirming that \textit{``many others share your feelings''} and proceeds to present others' similar experiences. Such validation helped participants feel ``held,'' reducing self-doubt and loneliness in vulnerable moments.

Participants described how Queerbot could also support them in exploring actionable guidance for navigating and performing their identities. 
% The chatbot was perceived as most helpful when it moved beyond reassurance and validation to also suggest potential pathways forward.
Participants appreciated that the chatbot surfaces supportive communities that were often hidden or overlooked on mainstream social media. They described these communities as being capable of providing essential social support and information about local LGBTQ+ movements and meet-ups. The chatbot also offered practical advice drawn from others’ experiences, ranging from making medical appointments to coping with gender-related challenges.

\begin{quote}
\textit{What I like the most is knowing that the path I want to take isn’t impossible, that someone has already tried it and shown it’s doable. That makes me more confident and reassured.} (P25).

\end{quote}

% This sense of support also extended to situations where participants felt unable or unwilling to respond directly to hostile discourse, but were reassured by knowing that Queerbot was on their side. In such cases, the chatbot’s replies in Ally Mode functioned as a proxy voice, offering relief through articulated responses from the user’s standpoint: \textit{``like a friend arguing back for me'' }(P18). Even if the original offender did not see the reply, simply viewing the words still provided comfort.

Such relief and assistance were most valuable during emotionally fragile times. Several participants described how receiving supportive resources or empathetic messages helped them feel less isolated about their own experience, especially when the topic is hard to share with real-life acquaintances, like family or friends. At the same time, participants recognized the limits of machine-generated comfort, with some finding it \textit{``a little hollow compared to real people''} (P23). For them, the most meaningful relief came when the chatbot pointed toward real-life lived experiences communities, rather than simulated empathy.

\subsubsection{Relating with diverse perspectives and lived experiences}

Participants valued Queerbot’s potential to broaden their outlook and help them situate knowledge within a wider range of perspectives. For example, several participants highlighted the desire to incorporate LGBTQ+-related discourses from different parts of the world and different points in history into the discussions with Queerbot. These examples serve as interpretive resources that make them aware of how issues could unfold differently depending on context. Such perspectives reminded them that what might be taken for granted locally in Taiwan, like views on gender roles or family, are far from universal, prompting further scrutiny of the status quo.

Participants also recognized the value of Queerbot in assisting them to provide empathetic responses to commenters with differing views, particularly when those individuals were well-intentioned but lacked adequate knowledge about gender and sexuality. For example, one stimulus material in Ally Mode featured a man struggling emotionally after his fiancée came out as asexual, asking whether it was simply an excuse for her lack of attraction toward him. Reflecting on this case, one participant noted: \textit{``I can clearly tell he needs emotional support, but he also desperately needs accurate knowledge about asexuality''} (P27). In such situations, Queerbot was seen as a bridge, offering both information and emotional framing to support constructive dialogue when users chose to engage.
% Even when disagreeing with another, participants valued being able to put themselves in another’s shoes: \textit{``even if the stance is different, I can at least see why they think that way''} (P26). 
For some, these moments also became opportunities to reflect on their own reasoning, to spot inconsistencies in their arguments, and to reconsider how they might position themselves. 
% Taken together, these experiences reveal how participants sought not only to defend their positions but also to engage dynamically with the environment to provide support, empathy, and foster critical thoughts.

\subsubsection{Being knowledgeable and rational in a discourse}

Another important dimension of benefit was Queerbot’s emphasis on providing and referencing the informative user-curated resources in responses. Several participants emphasized that their responses were valuable because they carried a basis in scholarship and gender theories, enabling them to explore further into an unfamiliar topic and be more knowledgeable in the discourse. This academic grounding was not only reassuring but also empowering: it equipped them with knowledge to comprehend and gave them greater confidence to participate in complex debates, such as surrogacy or gender-neutral restrooms. 

Furthermore, participants described the chatbot as a tool for helping them maintain a knowledgeable and rational character in communication. In Ally Mode, Queerbot generates different ways of responding to an inconsiderate response. This was described as a form of relief, since it allowed them to contribute to rational discussion with persuasive information at their disposal. Several also find value in using the chatbot to point out potential weak spots in their arguments, preventing their speech from being invalidated by others.

\begin{quote}
\textit{When I’m dealing with topics I’m not familiar with, I might need some external source to back up my opinion. If I already have an argument, I’d want to check with it, like, `Can you polish this for me, or point out any gaps?' ... In those situations, the chatbot can help you think ahead and detect parts of your argument that others might challenge.} (P26)
\end{quote}

% These results reveal participants’ latent desire to be more knowledgeably equipped in public discourse, and point to the potential of human-AI collaboration that leverages user-curated hermeneutical resources to support LGBTQ+ people's autonomy in discourse engagement.

\subsubsection{Balancing the debate and raising visibility}

The motivation to respond was not only about defending oneself but also about shaping the balance of the broader opinions. Several participants emphasized that replying to hostile or misinformed posts was less about persuading the original author and more about reaching other readers who might be influenced by harmful or discriminatory narratives. Being a member of the trans community, P21 described her motivation responding to transphobic speech: \textit{``maybe I can save an innocent teenager from becoming a TERF after seeing this post'' }(P21). Others described moments when Queerbot’s support helped them strengthen the visibility of minority perspectives that otherwise risked being buried by platform algorithms or hostile backlash.

At the same time, some participants also wish the chatbot could determine if they should engage in a discussion, considering the emotional labor involved. Hostile or dismissive posts were seen as draining and not worth the effort, while comments that came from genuine trauma or confusion felt more worthy of engagement: \textit{``He looks more hurt than malicious... so I think I can respond with empathy while also voice my stance''} (P21). Queerbot was therefore positioned as a tool not only for amplifying one’s opinion, but also for selectively channeling energy toward moments where the effort could matter most.

Finally, some participants underscored that asserting one’s stance did not always require friendliness. Effective visibility sometimes demanded a sharper tone, emotional force, or simply standing firm on a perspective. As one reflected, \textit{``it doesn’t have to be friendly to be educational''} (P22). By offering response options with different tones, Queerbot supported users in calibrating how they wished to be seen, both in defending their own position and in shaping the discourse around them.

% \subsubsection{Trust and concern for the platform and sources}

% Participants indicated that Queerbot’s usefulness depended on the integrity of the platform, and the credibility of its sources. Many expressed a cautious stance toward AI-generated claims, as one put it, \textit{“the first thing I do isn’t to read what the bot wrote, but to click the source and see what’s actually there”} (P18). One participant described frustration at generic invocations of “queer theory” without clear attribution: \textit{“Really? Who said that? How? Why?”} (P21). They point out that academic traditions like queer theory, which are frequently referenced by the chatbot, themselves contain multiple, often conflicting perspectives. In such cases, they are more reserved on chatbot's interpretations of the theory, and preferred to rely on the original sources, like books or articles, and build their understanding around it.

\subsubsection{Governing hermeneutical resources for the community}

Lastly, participants appreciated the value of LGBTQ+-centered design and curation of social support tools like Queerbot in addressing gaps in mainstream social platforms. Many saw value in Queerbot for harboring community members' experiences and providing interpretive tools otherwise unavailable. 

\begin{quote}
    \textit{So many times we have things to say, but no place to say them. Queerbot can be a bridge, a place where experiences are collected and recognized. Here you can use others’ life experiences or the concepts they’ve shared as a basis of comparison with your own, which then becomes a kind of bridge for further dialogue. Even if it doesn’t necessarily put you directly into a community, it brings you to a space where you can speak safely.} (P17). 
\end{quote}

% At the same time, community autonomy in the governance of such hermeneutical resources was seen as crucial. 
We observe that participants wanted an epistemic infrastructure curated by and for the community: \textit{``We trust because it’s built with queer people in mind, with sources that are supportive of diversity''} (P27). Others cautioned that without such control, spaces risked being invaded by ill-intentioned parties and becoming hostile or dangerous. This concern was amplified when considering individuals in vulnerable or exploratory stages of identity development: \textit{``if someone's identity is still very unstable or questioning, it’s not realistic to expect them to fact-check or critique claims made by a chatbot''} (P18). 

This dual emphasis on repairing interpretive gaps as well as ensuring self-determined control of hermeneutical resources is reflective of the participants' desire for a healthier epistemic infrastructure, as well as hermeneutical autonomy. 

\section{Discussion}
Our study investigated how Taiwanese LGBTQ+ people encounter hermeneutical injustice on social media and how technology might support hermeneutical autonomy in these environments. Through interviews, design workshops, and evaluation sessions, we revealed both the fragility and creativity of LGBTQ+ people’s strategies for navigating epistemic disadvantage, and we explored how a social support chatbot might scaffold more just epistemic conditions. 

\subsection{Hermeneutical injustice on social media for LGBTQ+ people} 

Hermeneutical injustice arises when marginalized people lack access to interpretive resources necessary to make sense of their experiences \cite{Fricker-2007}. Epistemic infrastructures, such as social media, can enable or obstruct the formation, uptake, and application of hermeneutical resources by shaping their visibility, circulation, and usage \cite{MilanoPrunkl-2025}.

In our formative study, we found that online communities played a pivotal role in situating Taiwanese LGBTQ+ people's identities. They often relied on online forums and communities to acquire concepts that helped them understand and articulate their sexual orientation or gender identity. Others drew on more structured sources, such as psychology articles or queer theory, to reframe their lived experiences with greater clarity. This complementarity echoes Queer HCI accounts of online environments affording identity work \cite{CraigMcInroy-2014, Cavalcante-2016, CarrascoKerne-2018, Haimson-2018, DymEtAl-2019, Wuest-2014, HaimsonEtAl-2020, FisherEtAl-2024} as well as narrative production and sharing \cite{HaimsonEtAl-2020, HardyEtAl-2022, SteedsEtAl-2025, TaylorBruckman-2024, ModiEtAl-2025, CuiEtAl-2022} for LGBTQ+ people, while extending them beyond single-community cases to account for their roles outside of the community \cite{TaylorEtAl-2024a,Haimson-2018}. It also resonates with findings that LGBTQ+ communities actively negotiate platform logics, assembling cultural toolkits to make themselves legible on their own terms \cite{DeVitoEtAl-2021a, TaylorBruckman-2024}.

The most consequential findings in our data concerned not the existence of hermeneutical resources, but their \textit{in situ} uptake and application. Supportive narratives were often algorithmically buried or socially drowned out in hostile threads, producing moments where participants felt they had “no place” to locate resonant counter-speech. This pattern converges with work on epistemic fragmentation and profiling that isolate users and hinder comparison and learning across experiences \cite{MilanoEtAl-2021,MilanoPrunkl-2025}. It also complements prior CHI/CSCW reports on platform causes frictions in areas including safety features, visibility incentives, and interface logics for LGBTQ+ users by showing how reductive, label-rewarding dynamics undercut nuanced self-presentation \cite{DeVitoEtAl-2021a,HaimsonEtAl-2020a}.

In our subsequent design workshops and evaluation workshops, participants’ appreciation of descriptive self-statements illustrates how interpretive resources can restore space for nuanced self-definition, countering the reductive logic of mainstream platforms that reward simplified labels. By reframing dismissive encounters, directing attention to supportive communities, or co-constructing knowledgeable responses to inconsiderate comments, participants reclaimed interpretive agency that was originally hindered by the failure of social media in undertaking the role of epistemic infrastructure \cite{MilanoPrunkl-2025}. These tactics refine earlier observations that identity-affirming sense-making is sustained through both safe spaces and outward-facing, precarious engagements \cite{DymEtAl-2019, Lucero-2017, CarrascoKerne-2018}. Relative to studies that prioritize insulated or identity-specific venues, our findings underscore that hermeneutical justice and resources must \textit{travel} with users into mixed or hostile arenas where resources are hardest to surface yet most needed. Situating these dynamics in Taiwan also adds nuance to regional and cultural scholarship. Participants calibrated visibility and retreat against a backdrop of legal–social mismatches and familist expectations (e.g., filial piety, compulsory marriage), shaping when and how they engaged \cite{Au-2022,Yeh-2014,Jhang-2020}. 

{\color{blue}
On a social-technical level, our findings illustrate a reciprocal relationship in which participants both shape and are shaped by the sociotechnical environment they inhabit. This dynamic resonates with Bardzell and Bardzell’s \cite{BardzellBardzell-2015} notion of the \textit{subjectivity of information}, wherein users can be simultaneously \textit{subjected to} the influence of certain sociotechnical environment, and exhibit agency as a \textit{subject of} those environment \cite{BardzellBardzell-2015}. 
In this context, our participants’ struggles with hermeneutical injustice can be understood as moments in which they are subject to the values inscribed on the epistemic infrastructure they engage with, and thus distorted their capacity to meaningfully make sense of their own experiences as a subject of the sociotechnical environment.

Our findings further foreground the stakes of reconfiguring both information infrastructures and the subjectivities they afford for LGBTQ+ people, reflected by our participants' desire for tools that expand rather than contract their epistemic horizons, allowing them to make sense of their own experiences on their own terms. The call emerging from our study is therefore not to \textit{support} hermeneutical work, but to \textit{reshape} the hermeneutical dynamics between users and infrastructure. From this, we explore a redesign of epistemic infrastructure towards hermeneutical autonomy \cite{AjmaniEtAl-2025}, through which we aim to cultivate and transform human agency in ways desirable to LGBTQ+ people \cite{BardzellBardzell-2015}.}

% Rather than a mere deficit of resources, we observe a lack of reachable, actionable resources at the point of interaction, created by social media platforms’ failure to function effectively as epistemic infrastructure for LGBTQ+ people \cite{SafirEtAl-2025}.

\subsection{Supporting hermeneutical autonomy for LGBTQ+ people}

Hermeneutical autonomy denotes people’s capacity to determine, govern, and apply the interpretive resources through which their experiences become intelligible \cite{AjmaniEtAl-2025,Fricker-2007}. Our findings suggest that, beyond the long-standing HCI emphasis on building “safe spaces” \cite{Lucero-2017,HaimsonEtAl-2020a}, LGBTQ+ users also need support that travels with them outside dedicated venues into mixed or hostile arenas, so that resources remain reachable and actionable at the moment of interaction. In this sense, we frame hermeneutical autonomy as an \textit{in-the-wild} practice enabled by epistemic infrastructures that couple community formation with reliable uptake and application \cite{MilanoPrunkl-2025,MilanoEtAl-2021}.

% Our prototype supported hermeneutical autonomy by enabling participants to (1) determine how they were understood by the system, (2) interpret their own social behavior and thoughts, and (3) access hermeneutical resources that help them navigate immediate social encounters. These mechanisms mitigated epistemic fragmentation, allowing participants to compare and connect their own experiences with those of others, and helped them become more intelligible in discourse as they wished.

Across design and evaluation workshops of our prototype, participants valued two categories of complementary affordances. First, mechanisms to shape and reflect the system’s understanding of the self (e.g., descriptive self-statements and transparent user models) supported exploration of fluid identities, resonating with prior Queer HCI work on identity work and cultural toolkits \cite{Haimson-2018, DeVitoEtAl-2021a, TaylorBruckman-2024}. Second, reflective prompts and curated, citable resources helped people reframe dismissive encounters to reduce distress, and assemble reasons and evidence to engage in public debates (e.g., surrogacy, gender-neutral restrooms), aligning with infrastructure views that emphasize not only resource formation but also their context-sensitive application \cite{MilanoPrunkl-2025}. Our observations thus extend earlier reports that prioritize protected or community-specific spaces \cite{HardyEtAl-2022,HardyVargas-2019} by showing how scaffolds must operate at the edges of those spaces where platform logics often reward reductive labels.

Taken together, these practices exemplify hermeneutical autonomy as proactive and dynamic work: users compare experiences, recalibrate self-descriptions, and mobilize community knowledge \textit{in situ} \cite{AjmaniEtAl-2025}. This stands in contrast to epistemic fragmentation, where algorithmic sorting and over-personalization impede access to counter-speech and comparative cases \cite{MilanoEtAl-2021,MilanoPrunkl-2025}. By positioning our system as a collaborator rather than a proxy, we operationalize community knowledge (via retrieval and transparent user models) so that hermeneutical labor becomes more accessible and less taxing, echoing calls in HCI to support autonomy through value-sensitive, context-aware tooling \cite{HaimsonEtAl-2020a,DeVitoEtAl-2021a}.

% For system designers, this highlights the need to build infrastructures that empower users to access and apply interpretive resources. AI tools should also help users avoid epistemic fragmentation \cite{SafirEtAl-2025}, enable meaningful comparisons across diverse experiences and contexts.

\subsection{Community governance of epistemic infrastructure}

Our research also revealed the importance of community member participation in designing and governing epistemic infrastructures, aligning with prior discovery by \cite{AjmaniEtAl-2023, TaylorBruckman-2024}, but extends to the practices of accessing and applying the hermeneutical resources curated or created by members of LGBTQ+ communities. Our prototype incorporated inputs and curated resources in its design as well as its materials. Participants valued how our prototype incorporated community-curated resources, which effectively supported identity exploration, interpreting social experiences, and navigating challenging environments.

While hermeneutical autonomy is initially theorized as an individual property \cite{AjmaniEtAl-2025}, our findings highlight that such autonomy also emerges through collaboration with peers and communities, mediated by technologies. Participants envisioned our chatbot not as a replacement for peer support, but as a medium that scaffolds accessing and applying hermeneutical resources created or curated by others. These results demonstrate that hermeneutical autonomy emerges through practices of shared governance, where individuals and communities co-curate interpretive resources, safeguard their legitimacy, and participate in the decisions that influence the specific epistemic conditions related to their interests. Technologies like social support chatbots can play a critical role here, not by replacing peer support, but by scaffolding the collective production, circulation, and application of hermeneutical resources, echoing previous suggestions by scholars \cite{WangEtAl-2021b, WanEtAl-2023}. Usage of techniques like RAG and natural language can also enhance the transparency and explainability of the mechanisms of epistemic infrastructure, facilitating fair and empowering participation of community members \cite{WanEtAl-2023}. In this way, community governance ensures that epistemic infrastructures serve both as protective shelters and as platforms that project marginalized voices into the wider public sphere.

{
\color{blue}
\subsection{Risks of AI-powered agents for support}

% Our study highlights the potential for AI agents to transcend the role of passive information retrievers and function as active partners in facilitating hermeneutical autonomy. 

% Kirk et al. \cite{KirkEtAl-2025} responded to this phenomenon by calling for AI designs that supports users in navigating tensions between immediate comfort and longer-term personal growth, and that attends to basic human needs of autonomy, competence, and relatedness \cite{RyanDeci-2008, RyanSapp-2007, KirkEtAl-2025}. 

Our work echoes prior research in showing how retrieval-augmented generation can be mobilized to enhance value sensitivity and cultural relevance for marginalized identities and non-western contexts \cite{SultanaEtAl-2025, ChangEtAl-2024, SeoEtAl-2025}. However, using pre-trained LLMs as an important component in the sense-making infrastructure of identity still risks introducing its inherent social bias and epistemic injustice, which were frequently manifested in prior work \cite{BragazziEtAl-2023, ByrnesSpear-2023, KayEtAl-2025, MarkoEtAl-2025, MeiEtAl-2023, PetzelSowerby-2025}. Even with explicit efforts to de-bias AI systems, users remain exposed to cultural and linguistic prejudices that are baked into contemporary LLM training methods \cite{LinEtAl-2024}. Furthermore, recent studies have shown that LLMs remain prone to hallucination \cite{FarquharEtAl-2024, DahlEtAl-2024}, even when paired with RAG strategies \cite{HuangEtAl-2025}. In our case, this could lead to the system inadvertently inventing theories, non-existing resources, or non-factual information. When conveyed through an affirmative tone, such problematic responses may be hard for users to distinguish, especially for those in moments of crisis or vulnerability \cite{WangEtAl-2025a}. 
% Echoing past study on the quality AI-provided empathetic response \cite{JakeschEtAl-2019, Seitz-2024, HuangEtAl-2024a}, participants in our study also noticed some moments when responses felt ``hollow'', and thus failed to provide appropriate and authentic emotional support.
These risks and failures demonstrate how AI deployment in high-stakes contexts can inadvertently exacerbate the harms it is intended to mitigate, and also foreground the importance of critically examining how the introduction of AI reconfigures the negotiations between users and the broader sociotechnical environment.

In our efforts to support hermeneutical autonomy, we attempted to position Queerbot as a responsible medium that scaffolds users’ own sense-making and contextualizes social experiences with real-world community or authoritative resources, rather than offering purely artificial empathetic responses. Despite this, prior works have repeatedly reported concerns about long-term usage of social support AI eroding \cite{FuEtAl-2025, ZhangEtAl-2025} and fostering emotional dependence \cite{ManziniEtAl-2024, ZhangEtAl-2025}, especially when systems are optimized for platform engagement \cite{ZhangEtAl-2025}, and use models that tend to agree with users or indulge short-term emotional gratification \cite{FuEtAl-2025, KirkEtAl-2025}. This tension, arising from the conflict between platform operators' commercial incentives and users' need for self-determination, significantly compromises the integrity of human-AI engagement by restricting the users’ decisional and discursive space within the dialogue. Such conflict shall be addressed in future development of social support AI systems with a more responsible paradigm for human-AI interaction \cite{KirkEtAl-2025}.

Our participants also echoed that a community-driven AI platform may be vulnerable to incursions by hostile actors, which could undermine community narratives and cultures that require substantial care to sustain \cite{TaylorBruckman-2024, AjmaniEtAl-2023, AjmaniEtAl-2024}. Additionally, since any hermeneutical resource will inevitably foreground some localities, concepts, and political agendas over others, Queerbot’s efforts to make certain identities legible may privilege more common lived experiences while rendering others less visible, potentially exacerbating existing intra-community tensions and the alienation of intersectional identities \cite{ModiEtAl-2025, TaylorBruckman-2024}. We encourage designers and researches to understand the nuance of these community dynamics before deploying similar community-driven AI systems.

\subsection{Design implications}

Our findings suggest that epistemic infrastructures, whether instantiated through social media, social support chatbots, or hybrid systems, must move beyond passive information delivery toward actively scaffolding hermeneutical autonomy \cite{Fricker-2007, Crerar-2016, AjmaniEtAl-2024}. This aligns with calls for \textit{socioaffective} alignment in social AI, which argue that systems should help users navigate tensions between short-term emotional comfort and longer-term growth in autonomy, competence, and relatedness, rather than simply maximizing engagement or agreement \cite{KirkEtAl-2025, ManziniEtAl-2024, RyanDeci-2008}. In line with prior work on social support chatbots \cite{Abd-AlrazaqEtAl-2020, WangEtAl-2021b}, our findings suggest that social support agents should not be framed as endlessly affirming “friends” that operate outside of users’ broader epistemic context, and point towards a more responsible design that incorporates well-informed decisions while providing companionship.

We identify several design directions for implementing systems that are situated within users’ social environments and draw on relevant hermeneutical resources when providing assistance. First, support agents can be implemented as dedicated conversational partners within messaging platforms (e.g. \cite{LiuEtAl-2024}), where users can forward challenging social content and receive guidance within the same space. Second, agents can be incorporated directly into social media contexts (e.g. \cite{SalehzadehNiksiratEtAl-2023}), allowing them to surface support in response to ongoing interactions in a more proactive manner. Third, embedded agents in browsers (e.g. \cite{DrouinEtAl-2024}) or augmented reality (AR) devices (e.g. \cite{YangEtAl-2025}) can access a user’s broader social context and the wider set of resources available across the web or even in physical environments; by leveraging new interface designs, such agents may also provide more fluent user experiences \cite{YangEtAl-2025}. While these possibilities can enable richer social context and more proactive support, we caution that such implementations must respect user self determination, and be aware of the risks mentioned in 6.4. 
% As Hopster \cite{Hopster-2024} warned, technologies that alter or disrupts existing social conditions can either strengthen or undermine users hermeneutically. Thus, by adding new layers of interpretation into online interactions, support agents may allow unfair or hostile intentions to penetrate more deeply into the platform logic. Designers should therefore ensure that system interventions remain sensitive to users’ own meaning making processes and respect user agency.

We've also shown how to operationalize hermeneutical autonomy on a system design level. AI systems should incorporate transparent, controllable user understandings encoded in natural language, enabling autonomy in how the system interprets a user and infer their social preferences accordingly. Such design of epistemic infrastructures that respects the self-determination of identity and social experience is essential in countering hermeneutical injustice exacerbated through algorithmic profiling and isolation \cite{SafirEtAl-2025, MilanoPrunkl-2025, MilanoEtAl-2021}. Effective epistemic support also requires systems to retrieve and surface hermeneutical resources that are both contextually relevant and diverse in form, from peer narratives to scholarly theories. Systems should therefore act as active interpreters that situates users’ encounters within broader cultural vocabularies and theoretical frameworks \cite{SultanaEtAl-2025, ChangEtAl-2024}. Systems can also leverage existing social media community contexts to help users navigate challenging social situations by accessing crucial information, as well as connecting with extant community members and discourse \cite{WangEtAl-2021b}.

Finally, our findings reinforce that epistemic infrastructures must be re-conceptualized as systems governed by the community, rather than as value-neutral products. To effectively empower community members in controlling the epistemic conditions utilized to interpret their experiences \cite{AjmaniEtAl-2025}, platforms should adopt frameworks of participatory governance \cite{AjmaniEtAl-2024} and distributed responsibility \cite{SultanaEtAl-2025} that validate non-traditional forms of localized knowledge. This commitment is also critical when resolving intra-community tensions and attending to intersectional nuances, which requires embracing heterogeneity by establishing minimal standards of shared belief rather than enforcing prescriptive definitions \cite{TaylorBruckman-2024}. To operationalize this, mechanisms can be designed to ensure inclusive community rules \cite{AjmaniEtAl-2024}, and communication between moderators and users shall be supported to ensure policies reflect specific cultural sentiments of marginalized sub-groups \cite{ModiEtAl-2025}.
% Prior CSCW and HCI research has shown that online LGBTQ+ communities, which are important venues for developing, sustaining, and circulating hermeneutical resources \cite{HaimsonEtAl-2020, HardyEtAl-2022, CraigMcInroy-2014, SteedsEtAl-2025, TaylorBruckman-2024, ModiEtAl-2025, CuiEtAl-2022}, invest substantial labor in building, curating, and defending their safe spaces. This body of work has also documented how epistemic injustice emerges when that labor or knowledge is dismissed, extracted, or undermined by inconsiderate or hostile actors \cite{AjmaniEtAl-2023, AjmaniEtAl-2024, TaylorBruckman-2024}. In our study, participants similarly expressed concern that a community driven chatbot could be infiltrated by antagonistic users or repurposed to police supposedly acceptable identities, reflecting long standing patterns in which queer and intersectional communities must continually defend the boundaries of their narrative spaces .

}

\subsection{Limitations and future directions}

Our prototype deliberately privileged a curated, text-forward corpus and conservative retrieval. This decision supported clearer provenance but narrowed the range of everyday sense-making artifacts participants actually encounter online (e.g., memes, screenshots with overlaid text, short-video clips). In hostile or fast-moving threads, these multimodal forms often carry salient cues for interpreting stance and intent. Future work can explore how multimodal retrieval and generation of non-text artifacts can affect the uptake and application of hermeneutical resources.

{\color{blue}Although our sample included a range of LGBTQ+ identities, most participants were university students in northern Taiwan. Research on student samples, however, suggests that they do not inherently threaten external validity when they are theoretically appropriate for the phenomena under study \cite{DruckmanKam-2011}. In our case, focusing on LGBTQ+ emerging adults is analytically fitting, as emerging adulthood (roughly ages 18–25) is a distinct life stage characterized by intensive exploration and integration of identities, including sexuality and gender \cite{Lopez-LeonCasanova-2025, Wagaman-2016, CraigMcInroy-2014, CraigEtAl-2020}. Nonetheless, future work should broaden the participant pool to include older, non-student, and rural Taiwanese LGBTQ+ people in order to better capture the diverse trajectories of queer lives.}

In our research, the abundance of results was limited by the absence of a field deployment, largely due to constraints in time and research resources. Future research could extend our findings by conducting longitudinal field studies to observe how LGBTQ+ people integrate such epistemic infrastructures into their everyday practices.

Finally, while our work focused on Taiwanese LGBTQ+ people broadly, future studies could attend more closely to differences across identities and explore intersectional dynamics. For example, indigenous LGBTQ+ individuals may negotiate distinct challenges shaped by indigenous cultural values instead of Confucian traditions. Attending to these variations would deepen understanding of how hermeneutical autonomy is constructed across diverse contexts.

\section{Conclusion}
In this work, we examined how Taiwanese LGBTQ+ individuals navigate hermeneutical injustice on social media and how technology might foster hermeneutical autonomy. We found that people rely on fragile yet creative strategies, such as seeking validation in comment sections, reframing hostility through interpretive resources, and leaning on allies, which sometimes fail because platforms underperform as epistemic infrastructure, impeding access to and application of relevant resources. To address these challenges, we designed and evaluated Queerbot, an LLM-powered chatbot that scaffolds identity exploration, surfaces interpretive resources, and supports conversation engagement. Our findings indicate that Queerbot can help reframe hostility, validate lived experiences, and broaden perspectives while assisting knowledgeable participation. Importantly, participants envisioned the chatbot as a collaborator rather than a proxy, extending community knowledge, amplifying marginalized voices, and supporting fluid self-exploration. {\color{blue}Taken together, these insights underscore that AI agents for social support must be designed as accountable, community-governed partners in hermeneutical work, not as value-neutral tools, while addressing concerns about emotional dependence, biased corpora, and fabricated or misleading responses that can further reproduce or amplify existing forms of epistemic and affective harm.} This work contributes to HCI by articulating hermeneutical justice as a design goal for AI systems and by demonstrating how retrieval-augmented, user-curated approaches can operationalize community knowledge in social contexts, while foregrounding the ethical obligations to surface provenance, mitigate hallucinations and bias, and preserve users’ self-determination. {\color{blue}We call for designers and researchers to build inclusive, community-controlled systems that enable AI and social platforms to properly uptake its role as epistemic infrastructure, reducing interpretive gaps and advancing hermeneutical autonomy and epistemic justice in digital environments.}

\begin{acks}
This work was made possible by the bravery and support from all the LGBTQ+ community members who participated, as well as years of enduring efforts of the Taiwanese LGBTQ+ movement that paved the way for the present inquiry. We honor their resilience and celebrate their contributions through this research. We also greatly appreciate the guidance and encouragement given by the reviewers in the revision process. We would also like to thank National Science and Technology Council (NSTC) and College of Management at National Taiwan University (NTU) for their financial support (113-2221-E-002-186-MY3 from NSTC and 115PPT703 from NTU). Finally, the first author wishes to dedicate this work to the memory of Sam and Moo-er, two dear friends who passed away during the revision of this work; their company and support will be forever remembered.
\end{acks}

\bibliographystyle{ACM-Reference-Format}
\bibliography{QUHI}

\appendix

\clearpage
\section{Appendix \#1. Interview Protocol for the Formative Study}
\subsection{Basic Information}
\begin{itemize}
    \item Could you briefly introduce yourself? (Please include your age, occupation, gender identity, and sexual orientation.)
\end{itemize}

\subsection{Social Media Practices}
\begin{itemize}
    \item Would you describe yourself as a frequent user of social media?
    \item Which social media platforms have you used in the past, and which ones do you currently use most often? (For example: Facebook, Twitter, Messenger, LINE.)
    \item On the platforms you have just mentioned, could you provide examples of groups, communities, or fan pages you have joined? With whom do you usually interact there, or what types of content do you typically view?
    \item Are there any groups or communities related to your LGBTQ+ identity? In those communities, do you present yourself most often in the role of a gay man?
    \item Among the social media platforms you use, as well as the groups, communities, and fan pages within them, which ones feel relatively ``safe, warm, and like-minded'' to you? Which ones feel more ``challenging'' or ``divergent in values''?
    \item When using these social media platforms (or engaging with these communities), what expectations do you have? Are these expectations generally met?
    \item What actions do you take to maintain your existing like-minded environment? For example: blocking, hiding, liking, sharing, hiding stories, adding to ``close friends,'' or creating/joining groups.
\end{itemize}

\subsection{Minority Stress and Emotion Regulation}
\begin{itemize}
    \item Could you share an experience in which you encountered negative or unfriendly comments related to the LGBTQ+ community on social media? When this occurred, what did you do to regulate your emotions?
    \item Which social media features did you use, or did you turn to specific communities or groups on these platforms? For example: ignoring, blocking, (retaliating), seeking support in other communities, or shifting to offline activities.
    \item What psychological or contextual factors influence the way you use social media for emotion regulation (e.g., your current emotional state, the closeness or distance of the source of the comments)?
    \item Could you share an experience in which you encountered negative or unfriendly comments related to the LGBTQ+ community in your daily life? When this occurred, what did you do to regulate your emotions?
    \item Which social media features did you use, or did you turn to specific communities or groups on these platforms? For example: ignoring, blocking, retaliating, seeking support in other communities, or shifting to offline activities.
    \item How did you learn and develop the emotion regulation strategies you have just described? Before establishing these strategies, how would you typically respond to negative or unfriendly comments related to the LGBTQ+ community?
    \item Could you share an experience in which, when attempting to seek emotional support or retreat into a like-minded environment, the outcome did not meet your expectations or even caused further harm?
    \item Following such an experience, how did you adjust the way you usually seek emotional support or comfort?
    \item Could you describe the impact that the 2018 Taiwan same-sex marriage referendum or the various anti-LGBTQ+ campaigns during that period had on your like-minded communities?
    \item During that period, how did you adjust the way you usually seek emotional support or comfort?
    \item Could you share an example of a time when you intentionally explored content outside your like-minded environment? What circumstances or psychological states prompted you to do so? For example: “undercover” participation in a Facebook group of an opposing political party or organization.
\end{itemize}

\subsection{Self-Reflection Behaviors and Design Considerations}
\begin{itemize}
    \item Have you ever engaged in self-reflection or self-assessment regarding your social media usage and emotion regulation strategies? For example, using time review features or ``Screen Time.'' For instance, feeling that at a certain point you should have been more proactive in seeking support, or realizing that you had become overly comfortable within your comfort zone.
    \item What kind of personal growth did such experiences bring you? How did they enhance your psychological resilience?
    \item Based on your past and current experiences, what suggestions would you offer to social media platforms to better support your emotion regulation strategies? For example, incorporating usage pattern analysis or mental health prompts.
\end{itemize}

\clearpage
\section{Appendix \#2. Themes and Codes from Formative Study}
\begin{table}[h!]
  \centering
  % \caption{Themes, Descriptions, and Codes}
  \label{tab:themes_codes}
  
  % Column widths:
  % 1. Theme: 15% (Enough for titles)
  % 2. Description: 30% (Enough for context)
  % 3. Codes: 50% (Widest for the long lists)
  % Total = 95% of textwidth (leaving 5% for padding)
  \begin{tabular}{ p{0.15\textwidth} p{0.30\textwidth} p{0.50\textwidth} }
    \toprule
    \textbf{Theme} & \textbf{Description} & \textbf{Codes} \\ 
    \midrule

    \textbf{Identity management} & 
    Strategies, goals, challenges, and considerations in identity-related sensemaking and disclosure. & 
    First notice of own sexual orientation, Coming out, Exploring own sexual orientation, Acquiring knowledge of sexual orientation, Finding people with same identity, Relatable experience, Disclosure of identity, Sense of LGBTQ+ identity on platform, Online alter-ego, Distancing self with the community, Disclosure of political stance, Allying with other LGBTQ+, Intra-group diversity, Hiding sexual identity, Rejecting labels 
    \\ \midrule

    \textbf{Coping with emotions} & 
    Strategies, mechanisms, challenges, and conceptualizations about coping with emotions induced by social experience. & 
    Leave the scene of argument, Time will fix everything, Unsatisfactory feedback from safe space, Finding comfort in comment sections, Lack of closure, Finding comfort from online retaliation, Sense of closure, Sharing with safe space, Expecting positive feedback, Seeking support, Ignore emotion, Friction in sharing, Lack of alternative strategy, IM private message, Offline meetup, Screenshot, Turning off social media, Private Discussion, Avoiding burdening others, Not expecting feedback, Finding emotional outlet, Processing own thoughts, Emotion absorbing 
    \\ \midrule

    \textbf{Appraisal of social experience} & 
    Strategies, goals, resources, challenges, and factors concerning the process of comprehending encountered social experiences. & 
    Non-issue because of non-close relation, Reaffirming own belief, Finding backup for own opinion, Literacy of opinion validity, Develop critical thinking, Academic training, Reasonable opposite opinion, Re-evaluating closeness of relationship, Strength of relationship, Uncertainty of hostility, Relating with opposite opinion, Ability to deconstruct online hostility, Toxic anonymous accounts, Knowing the background of aggressor, Sharing with friend, Sharing with partner, Closeness of relationship, Perception of hostile intention, Understanding of life experience, Current mental capacity, Finding comfort in comment sections, Lack of closure, Finding comfort from online retaliation, Evaluate support before argument, Seeking support, Avoiding burdening others, Safe because of lack of knowledge, Queer reading, Queer theory 
    \\ \midrule

    \textbf{Responding to opinions} & 
    Strategies, goals, challenges, and considerations when (not) responding to hostile or inconsiderate public opinions. & 
    Informative debate, Fight back, Losing an argument, Online debate, Unfriending, Reconciling with opposite opinion, Ineffective opinion exchange, Evaluate support before argument, Hermeneutic labor, Hiding comments 
    \\ \midrule

    \textbf{Type of discourse} & 
    Type of discussion and discourse faced/engaged by participants on social media. & 
    General public issue, Identity-related lived experience, Gender transition topic, Understanding public topic, Feminist topic, Gossiping, LGBTQ+ community issue 
    \\ \midrule

    \textbf{Mental impact} & 
    Types of mental impact from interacting with social media. & 
    Self-consuming, Feeling neglected, Feeling lonely, Feeling hurt 
    \\ \bottomrule

  \end{tabular}
\end{table}

\clearpage
\section{Appendix \#3. Scenarios Used During Design workshops}
\begin{enumerate}
\item \textit{Don't wanting to be labeled.}
\item \textit{Wanting to understand more about my own identities.}
\item \textit{Wanting to share about something, but can't find a proper space to do so.}
\item \textit{Having certain expectations and pressure from within the communities.}
\item \textit{Unable to find responses that support my stances in comment sections.}
\item \textit{Wanting to raise the visibility of a community or a topic.}
\item \textit{Feeling exhausted from repeatedly explaining about my identity. }
\item \textit{Struggling to interpret the
discomfort or stress experienced on social media}
\item \textit{Wanting to fight back, but worried about the communication being ineffective.}
\item \textit{Wanting to share something negative happened on social media, but worried about troubling others.}
\end{enumerate}

\clearpage
\section{Appendix \#4. Themes and Ideas from Design workshop }
\begin{table}[h!]
  \centering
  % \caption{Themes and Design Implications}
  \label{tab:themes}
  
  % We manually calculate widths to roughly equal \textwidth (approx 17cm or 7in).
  % 0.2 + 0.3 + 0.45 = 0.95 \textwidth (leaving 5% for padding/spacing)
  \begin{tabular}{p{0.2\textwidth} p{0.3\textwidth} p{0.45\textwidth}}
    \toprule
    \textbf{Theme} & \textbf{Description} & \textbf{Example Ideas} \\
    \midrule

    \textbf{Self} \par (Nuanced \& dynamic identities) & 
    Rich and contextual self-representation in social environment. & 
    % Note: Standard itemize has lots of padding. 
    % If allowed, use \usepackage{enumitem} and \begin{itemize}[nosep, leftmargin=*]
    \begin{itemize}
      \item When creating a profile, chat with me instead of filling forms.
      \item Personal profile that mixes labels with free-text description.
    \end{itemize}
    \\ \midrule

    \textbf{People} \par (Contextualized social connection) & 
    Connecting with people and communities of similar life experiences. & 
    \begin{itemize}
      \item After describing social experiences to a chatbot, recommend terms or groups.
      \item See other people’s stories and compare with your own.
      \item Export ``what we know about you'' as a dating-app profile.
    \end{itemize}
    \\ \midrule

    \textbf{Information} \par (Accessing resources) & 
    Accessing information that helps make sense of own life experience. & 
    \begin{itemize}
      \item Conversational UI that captures rich user intent and helps find information.
      \item A news feed that integrates with gender advocacy groups.
    \end{itemize}
    \\ \midrule

    \textbf{Environment} \par (Navigating challenges) & 
    Navigating challenging social environments and situations. & 
    \begin{itemize}
      \item Create an ally agent to handle repetitive communication.
      \item Display ``\textit{Let’s fight! Ally bot is ready!}'' after choosing to engage.
    \end{itemize}
    \\ 
    \bottomrule
  \end{tabular}
\end{table}

\clearpage
\section{Appendix \#5 Themes and Codes from Evaluation Workshops}
\begin{table}[h!]
    \centering
    
    \label{tab:themes_codes}   % Optional label

    % Define 3 columns. 
    % The widths (0.22 + 0.3 + 0.43 = 0.95) sum to approx \textwidth.
    % We use 'p' columns so text wraps automatically.
    \begin{tabular}{ p{0.22\textwidth} p{0.3\textwidth} p{0.43\textwidth} }
        \toprule
        \textbf{Theme} & \textbf{Code} & \textbf{Description} \\ 
        \midrule

        % --- Group 1 (2 rows) ---
        % We set the multirow width to be the same as the column width (=) or explicit size.
        % If your latex is old, replace {=} with {0.22\textwidth}
        \multirow{4}{=}{Making sense of my identity} 
        & Re-evaluation of self identity 
        & Reflecting on and reassessing one’s own identity in light of new experiences or insights. \\ 
        \cline{2-3} % Draw line only across columns 2 and 3
        
        & Asking relevant questions to help exploration 
        & Using guided questioning to support self-exploration and identity sensemaking. \\ 
        \midrule

        % --- Group 2 (4 rows) ---
        % Note: You have 4 items here, so we span roughly 8 lines of text depending on wrapping.
        % Sometimes just using the number of rows (4) is enough.
        \multirow{8}{=}{Accessing hermeneutic resources} 
        & Shared relief (feeling not alone) 
        & Experiencing emotional relief through realizing that others share similar feelings or experiences. \\ 
        \cline{2-3}

        & Exploration of community 
        & Actively seeking out and learning about relevant communities for belonging and understanding. \\ 
        \cline{2-3}

        & Exploration of identity performance 
        & Experimenting with how one presents or performs identity in different social contexts. \\ 
        \cline{2-3}

        & Relieving hermeneutic labor 
        & Reducing the cognitive and emotional effort required to interpret, justify, or explain one’s experiences. \\ 
        \midrule

        % --- Group 3 (3 rows) ---
        \multirow{6}{=}{Engaging with the environment}
        & Raising visibility of own identity or stance 
        & Making one’s identity or viewpoint more visible in public or social discussions. \\ 
        \cline{2-3}

        & Want to be more rational in a topic or discussion 
        & Striving to approach discussions with greater rationality, objectivity, or logical framing. \\ 
        \cline{2-3}

        & Want to relate with others in topic or discussion 
        & Desiring to connect with others emotionally or experientially within a shared discussion topic. \\ 
        \bottomrule

    \end{tabular}
\end{table}

\end{document}